\documentclass[twocolumns]{aa}
\usepackage{graphicx}
\usepackage{txfonts}
\begin{document}
%


\title{The synthesis of the cosmic X-ray background\\ 
in the Chandra and XMM-Newton era}

\author{
R. Gilli\inst{1}, 
A. Comastri\inst{1},
G. Hasinger\inst{2}
}

\authorrunning{R. Gilli et al.}  \titlerunning{The synthesis of the
cosmic X-ray background}

   \offprints{R. Gilli \\email:{\tt roberto.gilli@bo.astro.it}}

   \date{Received ... ; accepted ...}

\institute{
Istituto Nazionale di Astrofisica (INAF) - Osservatorio Astronomico di
Bologna, Via Ranzani 1, 40127 Bologna, Italy
\and
Max-Planck-Institut f\"ur extraterrestrische Physik, Postfach 1312,
D-85741 Garching, Germany
}

\abstract{We present a detailed and self-consistent modeling of the
cosmic X-ray background (XRB) based on the most up-to-date X-ray
luminosity functions (XLF) and evolution of Active Galactic Nuclei
(AGN). The large body of observational results collected by soft
(0.5-2 keV) and hard (2-10 keV) X-ray surveys are used to constrain at
best the properties of the Compton-thin AGN population and its
contribution to the XRB emission. The number ratio $R$ between
moderately obscured (Compton-thin) AGN and unobscured AGN is fixed by
the comparison between the soft and hard XLFs, which suggests that $R$
decreases from 4 at low luminosities to 1 at high luminosities.
From the same comparison there is no clear evidence of an evolution of
the obscured AGN fraction with redshift. The distribution of the
absorbing column densities in obscured AGN is determined by matching
the soft and hard source counts. A distribution rising towards larger
column densities is able to reproduce the soft and hard AGN counts
over about 6 dex in flux. The model also reproduces with excellent
accuracy the fraction of obscured objects in AGN samples selected at
different X-ray fluxes. The integrated emission of the Compton-thin
AGN population is found to underestimate the XRB flux at about 30 keV,
calling for an additional population of extremely obscured
(Compton-thick) AGN. Since the number of Compton-thick sources
required to fit the 30 keV XRB emission strongly depends on the
spectral templates assumed for unobscured and moderately obscured AGN,
we explored the effects of varying the spectral templates. In
particular, in addition to the column density distribution, we also
considered a distribution in the intrinsic powerlaw spectral indices
of variable width. In our baseline model a Gaussian distribution of
photon indices with mean $\langle \Gamma \rangle=1.9$ and dispersion
$\sigma_{\Gamma}=0.2$ is assumed. This increases the contribution of
the Compton-thin AGN population to the 30 keV XRB intensity by $\sim
30\%$ with respect to the case of null dispersion (i.e. a single
primary AGN powerlaw with $\Gamma=1.9$) but is not sufficient to match
the 30 keV XRB emission. Therefore a population of heavily obscured
-Compton-thick- AGN, as large as that of moderately obscured AGN, is
required to fit the residual background emission. Remarkably, the
fractions of Compton-thick AGN observed in the Chandra Deep Field
South and in the first INTEGRAL and Swift catalogs of AGN selected
above 10 keV are in excellent agreement with the model predictions.

}

   \maketitle

\section{Introduction} \label{introduction}

After more than 40 years since its discovery (Giacconi et al. 1962),
it is by now clear that the cosmic X-ray background (XRB) is produced
by the integrated emission of faint extragalactic point--like
sources. In the deepest X-ray surveys to date, i.e. the 800 ksec
XMM-Newton Lockman Hole (Hasinger 2004), the 1 Msec Chandra Deep Field
South (CDFS, Giacconi et al. 2002) and the 2 Msec Chandra Deep Field
North (CDFN, Alexander et al. 2003), the summed X-ray fluxes of the
detected sources account for virtually the entire XRB emission around
a few keV and for about half of that at higher (7--10 keV) energies
(Worsley et al. 2005), the exact fraction depending on the absolute
XRB flux which is still affected by rather large uncertainties. The
measurements performed by the HEAO1--A2 experiment (Marshall et
al. 1980), which provided the only broad band (3--100 keV) XRB
spectrum available to date, has the lowest normalization. More recent
values, obtained below 10 keV with imaging instruments, were found to
be always higher, the highest one having been measured by BeppoSAX
(Vecchi et al. 1999). The most recent determinations of the XRB flux
by RXTE (Revnivtsev et al. 2003), XMM-Newton (Lumb et al. 2002; De
Luca \& Molendi 2004) and {\it Chandra} (Hickox \& Markevitch 2006) as
well as a reanalysis of the HEAO1--A2 data (Revnivtsev et al. 2005)
have been found to lay in between the original HEAO1--A2 measure and
the BeppoSAX one, comparable to those obtained a few years ago by ASCA
(Gendreau et al. 1995; Kushino et al. 2002).  It is worth mentioning
that, on the basis of a careful determination of the unresolved X--ray
background in {\it Chandra} deep fields, Hickox \& Markevitch (2006)
concluded that the contribution of resolved sources accounts for $\sim
80\%$ of the 0.5--8 keV {\it Chandra} flux which, in the 2--8 keV
band, corresponds to the entire HEAO1--A2 flux.  At energies $>$ 10
keV, where the bulk of the XRB energy resides, the only available
measure is that of HEAO1. Whether the high energy flux should be
renormalized upwards by $\sim 30\%$ (see e.g. Ueda et al. 2003) or the
very shape of the broad band spectrum is different from that
originally observed is still the subject of debate (see Gilli 2004 and
Revnivtsev et al. 2005 for recent discussions). The issue of the XRB
intensity, especially around its 30 keV peak, bears important
consequences for the energy released through accretion onto
supermassive black holes integrated over the cosmic history. Indeed,
extensive optical follow--up programs have shown that the vast
majority of the XRB sources are identified as Active Galactic Nuclei
(AGN) with different degrees of obscuration (e.g. Szokoly et al. 2004,
Barger et al. 2005) confirming early predictions by AGN synthesis
models (Setti \& Woltjer 1989, Comastri et al. 1995). Among the many
issues which are currently subject of intense research activity there
are: a complete census of the entire AGN population and especially of
the most obscured objects; the space density of obscured sources as a
function of redshift and intrinsic X--ray luminosity; the demography
of high redshift ($z>$ 3) quasars; the black hole mass function; the
average radiative efficiency and Eddington ratios of accreting
supermassive black holes.


In order to properly address the above described scientific goals, the
most efficient approach is to combine deep pencil beam {\it Chandra}
and XMM--{\it Newton} pointings with shallower surveys over much
larger sky areas (e.g. the ASCA Medium and Large Sensitivity Surveys;
Ueda et al. 2001; the {\rm HELLAS} and {\rm HELLAS2XMM} surveys; Fiore
et al. 2001, 2003). The extensive multiwavelength coverage available,
though at different levels of sensitivity for the various X--ray
surveys, allows to sample the X--ray sky source content over several
orders of magnitude in the X--ray flux and has already provided
statistically meaningful (several hundreds) samples of AGN across a
wide range of redshifts, luminosities and obscuring column densities.

The redshift distribution of AGN selected both in the soft (0.5--2
keV) and hard (2--10 keV) X--ray band peaks at $z\simeq$ 0.7--1.0 and
is dominated by low luminosity objects (log L$_X$=42--44 erg
s$^{-1}$). The observed distribution is in contrast with the
predictions by simple models based on pure luminosity evolution (PLE)
or pure density evolution (PDE) of AGN through cosmic times, and has
led to a more complicated scenario. Indeed, the evolution of the
X--ray selected AGN luminosity function up to redshifts of $\sim$ 3--4
is described, both in the soft (Hasinger, Miyaji \& Schmidt 2005) and
hard (Ueda et al. 2003, La Franca et al. 2005, Barger et al. 2005)
X--ray bands, by a luminosity dependent density evolution model (LDDE)
whose best fit parameters imply a strong dependence of the AGN space
density and evolution with X--ray luminosity. More specifically, a
clear trend for an increase of the peak redshift for increasing X--ray
luminosity of the AGN space density is present. The comoving space
density of luminous (log $L_X > 44$) quasars peaks at $z \simeq$
2--2.5 and declines rapidly, by more than two orders of magnitude,
towards $z=0$, in marked contrast with the behavior of lower
luminosity AGN, whose space density increases by less than a factor 10
from $z=0$ to $z\simeq$ 0.7--1 and decreases thereafter.

Given that luminous quasars are on average expected to be powered by
more massive black holes than lower luminosity AGN, the luminosity
dependent evolution of AGN space density bears some similarities with
the evolution of the star formation history and mass assembly of
non-active galaxies and is often considered as a strong evidence of
the co-evolution of supermassive black holes and their host galaxies.
The cosmic downsizing of the AGN population can be explained assuming
that rare, massive black holes powering high redshift luminous quasars
are quickly formed and efficiently fed at early cosmic times in a gas
and dust rich environment, which makes possible accretion at high
Eddington ratios. The bulk of lower luminosity AGN population has to
wait much longer to grow and/or to be activated, possibly due to the
lack of large reservoir of gas which limits the accretion to a
sub-Eddington regime (see e.g Hopkins et al. 2006 and references
therein).


According to the above described theoretical framework a large
population of obscured high redshift quasars is predicted.  While it
has been proposed that they could be associated with high-redshift
submm sources (Alexander et al. 2005), the actual size of the obscured
QSO population at high redshift is however still largely debated. The
most recent results from the largest compilations of X-ray surveys and
X-ray background synthesis models provide contradictory results,
sometimes favoring an increase of the obscured AGN fraction with
redshift (e.g. La Franca et al. 2005, Ballantyne et al. 2006),
sometimes favoring a constant fraction (Ueda et al. 2003, Treister \&
Urry 2005). On the contrary, evidence for a dependence of the fraction
of obscured AGN on the intrinsic source luminosity is rapidly growing,
the fraction of obscured QSOs being significantly lower than the
corresponding fraction of obscured Seyferts (Ueda et al. 2003,
Hasinger 2004, La Franca et al. 2005, Georgakakis et al. 2006).

Another still open and challenging issue, strictly related with that
of obscured high redshift quasars, concerns the number density of
heavily obscured, Compton-thick sources. Although X--ray observations
are the least biased against obscured AGN, nonetheless sensitive
imaging surveys can be performed at present only at energies below 10
keV and are therefore expected to severely undersample the number of
extremely obscured AGN with column densities above log$N_H=24$
(i.e. Compton-thick sources). In the local Universe Compton-thick AGN
are found to be as numerous as moderately obscured AGN (Risaliti et
al. 1999; Guainazzi et al. 2005) and are expected to provide a
significant contribution to the 30 keV peak in the XRB spectrum
(Comastri et al. 1995; Gilli et al. 2001). A full, self-consistent
modeling of the XRB spectrum which takes into account all the
available observational constraints appears to be, at present, the
most effective way to characterize the population of heavily obscured
Compton--thick AGN and thus to have a complete map of the AGN
demography. The space density and redshift distribution of
Compton-thick AGN is particularly relevant to properly estimate the
supermassive black hole mass function and, by comparison with the
relic black hole mass distribution of local galaxies, to constrain the
average Eddington ratio and radiative efficiency averaged over the
cosmic times (Marconi et al. 2004).


Throughout this paper a flat cosmological model with
$\Omega_m=1-\Omega_{\Lambda}=0.3$, $H_0=70\,h_{70}$ km s$^{-1}$
Mpc$^{-1}$ is assumed.

\section{Model outline}\label{model}

The wealth of X--ray data collected by sensitive imaging surveys in
the 0.5--10 keV energy range, complemented by optical and near infrared
photometric and spectroscopic identification programs which have
reached a high degree of completeness, allows us to tightly constrain
the contribution of unobscured and Compton-thin AGN to the XRB. Once
the overall properties of these AGN are constrained at best of our
present knowledge (e.g. source counts, luminosity functions and
redshift distributions, resolved XRB fraction) it is possible to
compute their contribution to the 0.5--300 keV XRB emission, for
reasonable assumptions on their spectral shape above 10 keV, and then
estimate the abundance of Compton-thick AGN by adding as many sources
as required to fit the residual (i.e. not accounted for by unobscured
and Compton-thin AGN) XRB intensity.

Given the well known discrepancies among different XRB measurements,
which still have to be fully understood, we will not adopt the XRB
emission below 10 keV as a primary model constraint. Rather, we rely
on what we believe to be more robust observational data such as the
0.5-2 keV and 2-10 keV logN--logS and luminosity functions and verify
{\it a posteriori} the resulting XRB spectrum. As far as the absolute
intensity of the 30 keV peak is concerned we will adopt the only
available HEAO1--A2 measurement (Marshall et al. 1980) as a reference
value. We discuss in Section~\ref{xrbnorm} the implications of a
higher level of the $>$ 10 keV XRB

More specifically the adopted strategy is as follows:


The 0.5--2 keV type-1 AGN luminosity function recently computed by
Hasinger, Miyaji \& Schmidt (2005) is assumed as the XLF of X--ray
unobscured AGN (log$N_H<21$), while the 2--10 keV AGN luminosity
functions obtained by Ueda et al. (2003) and La Franca et al. (2005)
are assumed to represent the total XLF of unobscured plus Compton-thin
(log$N_H<24$) AGN.

The soft XLF of unobscured AGN is converted in the 2--10 keV band
assuming a distribution for the X--ray spectral slopes of unobscured AGN.
The average value and dispersion are chosen to match those 
observed for relatively bright X--ray selected type-1 AGN.

The difference between the observed 2--10 keV XLF and the one
determined for unobscured AGN provides the total number of obscured
Compton-thin AGN ($21<$log$N_H<24$). The comparison
between the XLFs is also used to verify if there are variations in the
ratio between obscured and unobscured AGN as a function of redshift
and/or luminosity.

Once the total number of obscured Compton-thin AGN is determined,
their absorption distribution is found by fitting the source counts in
the 0.5--2 keV, 2--10 keV and 5--10 keV bands. When the
absorption distribution is determined, the population of unobscured 
and Compton-thin AGN is fully characterized.

Assuming an accurate modeling of the X--ray spectra of Compton-thin
AGN over the 0.5 - 400 keV energy range, including X--ray absorption,
a reflection component and a high energy exponential cut--off, their
contribution to the broad band XRB spectrum is computed.

The number of Compton-thick sources is then determined by adding as
many sources as needed to fit the residual XRB intensity, in
particular the XRB peak at 30 keV.

At this stage the model is checked for self-consistency by recomputing
the source counts in the various bands and the fraction of obscured
AGN as a function of the X--ray flux, taking into account the
contribution of Compton-thick AGN. Finally, we compute the model
predictions beyond the present observational limits and in particular
the expected fraction of Compton-thick AGN as a function of the survey
depth and energy range (i.e. below and above about 10 keV).

\section{AGN X-ray spectra}

\subsection{Unobscured AGN}\label{unabs}

The average X--ray spectral properties of AGN are relatively well
known over a broad range of luminosities and redshifts. The main
spectral component is a power law with an exponential roll over at
high (above 100 keV) energies. The average slope of the power law
component as measured for large samples of low-luminosity Seyfert
galaxies (i.e. Nandra \& Pounds 1994) and bright QSOs (Reeves \&
Turner 2000, Piconcelli et al. 2005) is close to $\Gamma$=1.9 without
any significant redshift dependence up to $z\lesssim 5$ (Vignali et
al. 2005, Shemmer et al. 2006, Risaliti \& Elvis 2006). The
distribution of the power law indices around the mean is found to have
a significant dispersion ($\sigma_{\Gamma} \simeq$ 0.2--0.3; Mainieri
et al. 2002, Mateos et al. 2005, Tozzi et al. 2006) which has always
been neglected in previous synthesis models. We will extensively
discuss the effects of a distribution of power law slopes in
Section~\ref{disp}. The shape of the continuum spectrum above a few
tens of keV is only poorly determined due to the lack of sensitive
observations at energies $>$ 10--20 keV.  While it is well known that
an exponential cut--off must be present around a few hundreds of keV
in order not to violate the present level of the XRB above 100 keV,
its characteristic energy $E_c$, if any, has been measured only for
few nearby bright Seyfert galaxies (see e.g. Matt 2001 for a review of
the BeppoSAX observations). The most recent results indicate a
relatively large spread in $E_c$ from about 50 keV up to about 500 keV
(Perola et al. 2002, Molina et al. 2006). In the following we adopted
a value of 200 keV, which is well within the observed range. The
impact of a different choice will be addressed in the Discussion
section.

In addition to the primary powerlaw, a hardening of the spectrum above
8--10 keV, coupled with an iron emission line at 6.4 keV, is seen in
AGN of Seyfert-like luminosity. Both are usually ascribed to
reprocessing by a cold accretion disk of the primary power-law
emission (Nandra \& Pounds 1994). Following Comastri et al. (1995), we
assumed a relative normalization of 1.3 for the reflected component in
type-1 unobscured objects, which is appropriate for an angle averaged
face--on reflecting slab covering a solid angle of $2\pi$ at the
primary source. The shape of the reflection continuum as a function of
the slope of the illuminating power law is self-consistently computed
for the adopted distribution of spectral slopes (see
Section~\ref{disp}). Following Gilli et al. (1999a), the 6.4 keV iron
emission line was modeled by a broad Gaussian of width 0.4 keV and
equivalent width 280 eV. Recent estimates of the average iron line
profile in type-1 AGN from deep XMM--{\it Newton} (Streblyanska et
al. 2004) and {\it Chandra} (Brusa, Gilli \& Comastri 2005)
observations suggest the presence of a red wing.  The skewed profile
is in agreement with emission from a relativistic accretion
disk. Since the details of the iron line contribution to the XRB
spectrum are beyond the scope of this paper and have been discussed
elsewhere (e.g. Gilli et al. 1999a), we will simply adopt a Gaussian
line parametrization. The reflection and iron line components are not
included at high (QSO--like) luminosities since there are
observational evidences against their presence (e.g. Nandra et
al. 1997, Page et al. 2005).

\begin{figure}
\includegraphics[width=9cm]{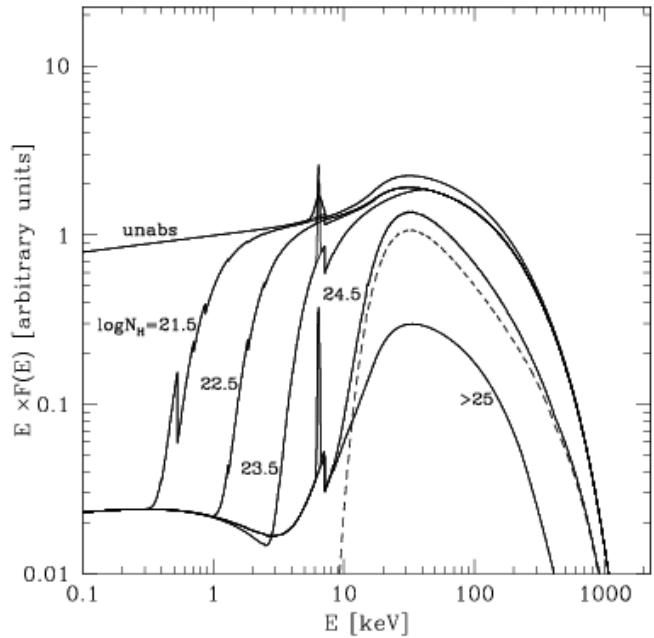}
\caption{The AGN X-ray spectra with different absorptions assumed in
the model. Solid lines from top to bottom: $N_H=0$ (i.e. unabsorbed
AGN), log$N_H=21.5,22.5,23.5,24.5,>25$. A primary powerlaw with
$\Gamma=1.9$ and cut off energy $E_c=200$ keV is assumed. A 3\% soft
scattered component is also added in the obscured AGN spectra. The
spectrum of mildly Compton-thick AGN (log$N_H=24.5$) is obtained by
summing a transmission component (dashed line) to the same reflection
continuum used to model the spectrum of heavily ($logN_H>25$)
Compton-thick AGN (see text).  In each spectrum a 6.4 keV iron
emission line is also added following Gilli et al. (1999a).}
\label{xspec}
\end{figure}

We note here that the spectrum of AGN in the soft band often shows a
higher degree of complexity with respect to that of a simple
powerlaw. On the one hand, observations of bright nearby Seyferts and
QSOs reveal a soft excess below 1--2 keV (Perola et al. 2002; Porquet
et al. 2004; Piconcelli et al. 2005) in many objects, which has been
interpreted either as the exponential tail of the thermal emission
from the accretion disk (Czerny \& Elvis 1987) or more recently as due
to a mildly ionized reflection component including a blend of emission
lines blurred by relativistic effects (Crummy et al. 2006). On the
other hand, absorption edges below about 1 keV due to ionized gas (the
warm absorber) are often present in a non negligible fraction of
nearby Seyfert galaxies and QSOs (Reynolds 1997, Porquet et al. 2004,
Piconcelli et al. 2005).  Given this complex situation, where some
objects show an excess and some others a deficit in the soft band, the
adopted spectral template of unobscured sources is approximated with a
single power law. The broad band 0.1--1000 keV spectrum of an
unobscured Seyfert with $\Gamma=1.9$ is shown in Fig.~\ref{xspec}.

We also note that, as far as the modelling of the broad band XRB spectrum is
concerned, any spectral complexity below about 1 keV has little impact
being shifted at very low energies by redshift integration.  It should
also be noted that a different prescription for the soft X--ray
spectrum and in particular the presence of a soft excess in the 0.5--2
keV band would affect the conversion factor with the 2--10 keV
band. We will investigate the effects of a different conversion factor
in the Discussion.

\subsection{Obscured AGN}\label{abs}

X--ray absorption by cold gas with column densities ranging from
values ($N_H\lesssim10^{21}$ cm$^{-2}$) barely detectable in the
X--ray band to extremely large columns able to efficiently absorb the
X--ray emission up to 10 keV ($N_H\gtrsim10^{24}$ cm$^{-2}$) are
routinely observed for both local (Risaliti et al. 1999) and distant
AGN (Norman et al. 2002, Tozzi et al. 2006, Mainieri et al. 2005).
For the sake of simplicity and following previous analysis (Comastri
et al. 1995, Gilli et al. 2001) the absorbing column density
distribution is parametrized with coarse, equally spaced, $\Delta
logN_H$ = 1 bins centered at log$N_H$=21.5,22.5,23.5,24.5,25.5.  The
slope of the primary continuum is that of unobscured AGN with a lower
reflection normalization (0.88 rather than 1.3).  Indeed, if the
accretion disk is aligned with the obscuring torus, the reflection
efficiency for high inclination angles (expected for obscured AGN) is
lower. A significant difference in the reflected component from the
accretion disk in unobscured and obscured AGN is not yet seen in the
data, however the typical values measured in samples of Seyfert 1s and
Seyfert 2s are consistent with the assumed values (Risaliti 2002,
Perola et al. 2002). As in the case of unobscured AGN, the disk
reflection component was added only for sources at Seyfert-like
luminosities.

The photoelectric absorption cut--off in Compton-thin AGN
(log$N_H<24$) is computed using the Morrison \& Mc Cammon (1983) cross
sections for solar abundances, while absorption in the Compton-thick
regime was computed with an upgraded version of a Monte Carlo code
originally developed by Yaqoob (1997). As long as the absorption
column density does not exceed values of the order of 10$^{25}$
cm$^{-2}$, the nuclear radiation above 10--15 keV is able to penetrate
the obscuring gas. For higher column densities the X--ray spectrum is
down-scattered by Compton recoil and hence depressed over the entire
energy range. In the following we will refer to sources in the
log$N_H=24.5$ and log$N_H=25.5$ absorption bins as ``mildly
Compton-thick''and ``heavily Compton-thick'' AGN, respectively.  The
broad band spectrum of mildly Compton-thick AGN is parametrized by two
components: the transmitted one which dominates above 10 keV (dashed
line in Fig.~\ref{xspec}) and the reflected one which emerges at lower
energies and is likely to be originated by reflection from the inner
side of the obscuring torus. As for the relative normalization of the
torus reflection component, which is only poorly known, we assumed a
value of 0.37. This is such to produce a 2--10 keV reflected flux
which is 2\% of the intrinsic one and is consistent with the average
ratio between the observed 2-10 keV luminosity of Compton-thick and
Compton-thin AGN (see e.g. Maiolino et al. 1998a). A pure reflection
continuum is assumed to be a good representation of the spectrum of
heavily Compton-thick AGN over the entire energy range. As for the 6.4
keV iron emission line, a Gaussian with $\sigma=0.1$ keV was assumed
for all obscured AGN, with equivalent width ranging from a few
hundreds eV for moderately obscured sources to $\sim 2$ keV for
Compton-thick objects (see details in Gilli et al. 1999a).


\begin{figure}[t]
\includegraphics[width=9cm]{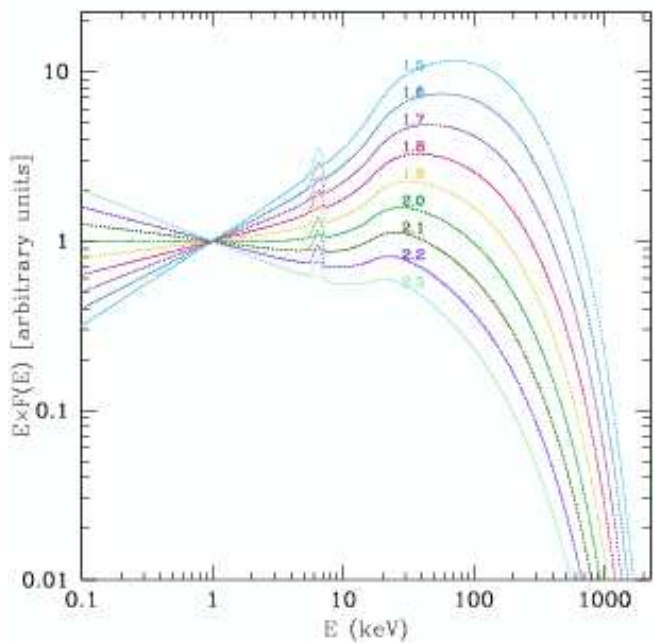}
\caption{The spectral library assumed for unobscured Seyferts. Photon
indices range from $\Gamma=1.5$ to $\Gamma=2.3$ in step of
$\Delta\Gamma=0.1$. All the spectra are normalized at 1 keV.}
\label{allgamma}
\end{figure}

\begin{figure}[t]
\includegraphics[width=9cm]{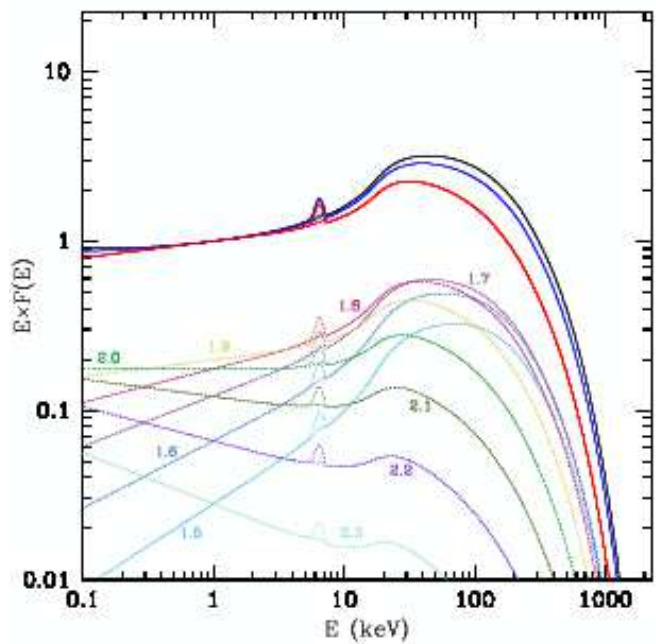}
\caption{The effects of increasing the spectral dispersion
$\sigma_{\Gamma}$ on the effective spectrum of unobscured AGN. The
solid lines normalized at 1 keV correspond to the weighted sum over
the assumed spectral distribution in the case of
$\sigma_{\Gamma}$=0.0, 0.2 and 0.3 (from bottom to top curve,
respectively). The case with null dispersion corresponds to assuming a
single spectrum with $\Gamma=1.9$. The contribution of the individual
spectra along the distribution (dotted lines with labeled photon
index) is also shown for the $\sigma_{\Gamma}=0.2$ case. Note that at
different energies the maximum contribution is provided by spectra
with different slopes.}
\label{xdisp}
\end{figure}

We finally added a soft component to the spectra of obscured AGN,
since soft X-ray emission in excess of the absorbed powerlaw is common
in the spectra of Seyfert 2 galaxies (e.g. Turner et al. 1997,
Guainazzi \& Bianchi 2006). The origin of this soft component has been
proposed to be manifold: i) a circumnuclear starburst, which is often
observed around the nuclei of Seyfert 2 galaxies (e.g. Maiolino et
al. 1998b) ; ii) scattering of the primary powerlaw by hot gas (Matt
et al. 1996); iii) leakage of a fraction of the primary radiation
through an absorbing medium which covers only partially the nuclear
source (Vignali et al. 1998, Malaguti et al. 1999); iv) sum of
unresolved emission lines from photoionized gas (Bianchi et
al. 2006a). Very recently Guainazzi \& Bianchi (2006) have shown that,
when observed with high spectral resolution, the soft X-ray emission
of most Seyfert 2 galaxies can be explained as the sum of individual
emission lines. However, since we are interested in the broad band
properties of the X-ray spectrum, we keep a simple modeling of the
soft X-ray spectrum by assuming a powerlaw with the same spectral
index of the primary powerlaw. Following Gilli et al. (2001), the
normalization of the scattered component was fixed to 3\% of that of
the primary powerlaw. This soft emission level is in agreement with
the recent results mentioned above (Guainazzi \& Bianchi 2006, see
also Bianchi et al. 2006b).

The broad band spectral templates adopted in the calculations
in the various absorption bins are shown in Fig.~\ref{xspec}.

\section{Dispersion of the photon indices}\label{disp}

While it is well known since the first AGN spectral surveys
(i.e. Maccacaro et al. 1988) that the distribution of X--ray spectral
slopes is characterized by a not negligible intrinsic dispersion, this
very observational fact has always been neglected in the synthesis of
XRB spectrum. The most recent estimates (Mateos et al. 2005) from the
analysis of relatively faint AGN in the Lockman Hole are consistent
with an intrinsic dispersion of about 0.2--0.3 confirming previous
results obtained for nearby bright Seyferts (Nandra \& Pounds 1994).
In the following we will discuss how the inclusion of a distribution
of spectral slopes modifies the synthesis of the XRB spectrum with
respect to the adoption of a single value.

It is easy to show that the spectrum obtained by averaging 
power law spectra with different slopes is different from that obtained 
computing the average slope :

\begin{equation}
\frac{1}{N}\sum_{i=1}^N\;E^{-\Gamma_i} \neq E^{-\langle\Gamma\rangle}.
\end{equation}

We computed a set of unabsorbed AGN spectra with photon indices
ranging from $\Gamma$=1.5 to 2.3, in step of $\Delta\Gamma=0.1$ to
which we added the previously described (Section~\ref{unabs}) spectral
components.  All the spectra are normalized at 1 keV and are shown in
Fig.~\ref{allgamma}. We assumed that the spectral distribution is a
Gaussian centered at $\Gamma=1.9$ and variable dispersion
$\sigma_{\Gamma}$. Each spectrum with photon index $\Gamma_i$ was
multiplied by the appropriate Gaussian weight $p_i$, i.e.:

\begin{equation}
p_i = \frac{1}{\sigma_{\Gamma}\sqrt{2\pi}}\int_{\Gamma_i-\Delta\Gamma/2}^{\Gamma_i+\Delta\Gamma/2}
e^{-(\Gamma-\langle\Gamma\rangle)^2/2\sigma_{\Gamma}^2}{\rm d}\Gamma,
\label{pii}
\end{equation}

with 
\begin{equation}
\sum_{i=1}^N\,p_i = 1.
\label{sumpi}
\end{equation}

All the weigthed spectra were then summed together to produce an
``effective'' spectrum, which is shown in Fig.~\ref{xdisp} as a
function of the assumed dispersion $\sigma_{\Gamma}$.\footnote{
Since we are considering a spectral range limited to $1.5<\Gamma<2.3$,
the sum in Eq.~\ref{sumpi} is slightly less than 1 (say by a factor
$f$) if the spectral dispersion $\sigma_{\Gamma}\gtrsim 0.2$. In this
case we simply scale the $p_i$ weights up by $f$ to make
Eq.~\ref{sumpi} hold.}

The slope of the resulting spectrum hardens towards high energies for
increasing values of the dispersion. By definition a null dispersion
corresponds to a single power law with $\Gamma=1.9$. A dispersion of
0.2-0.3 produces an increase by 30-40\% at energies above $\sim 30$
keV. Indeed, the relative contribution of the flatter slopes is higher
at high energies, while the relative contribution of the steeper
slopes is correspondingly increasing below 1 keV. As we will see
later, the spectral hardening at high energies and especially around
30 keV due to the adopted dispersion has a significant impact on the
XRB fit, affecting in turn the number of heavily obscured sources
required to match the high energy XRB spectrum.

In addition to the distribution in the spectral indices we also
considered the effects of a Gaussian distribution of the cut--off
energy values, centered at $E_c=200$ keV and with variable dispersion
encompassing the 50--500 keV energy range. 
The differences with respect to a single spectrum with zero dispersion 
are found to be very small, unless an extremely large dispersion
corresponding to an almost flat distribution, is assumed. 
Even in that case, however, the deviations with respect to 
the zero dispersion case are observable only above 100--150 keV, and thus do
not significantly alter our 30 keV XRB modeling. In the following we
will therefore consider a single cut--off energy $E_c=200$ keV.


\begin{figure*}
\includegraphics[width=18cm]{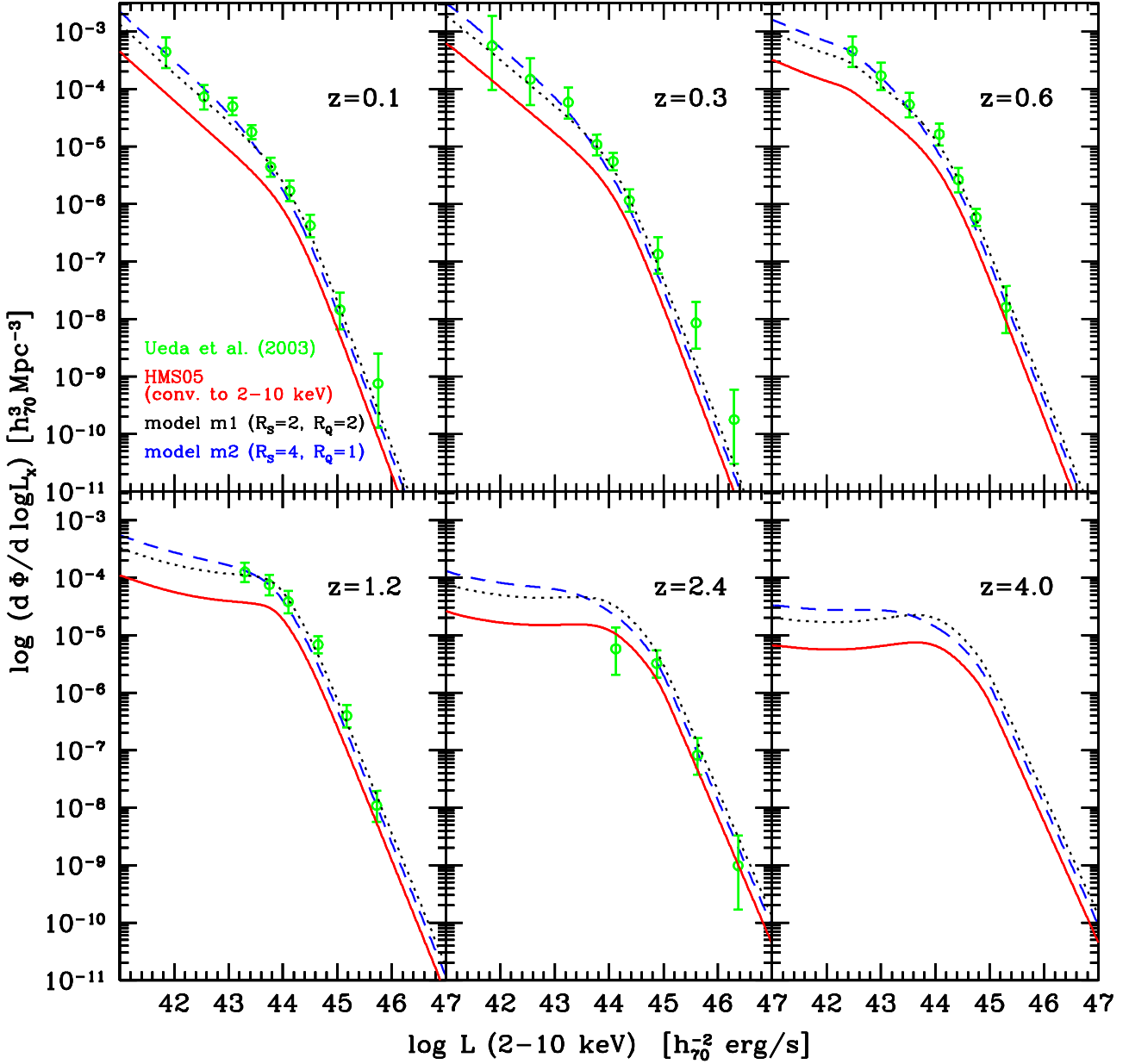}
\caption{The Ueda et al. (2003) intrinsic hard XLF for the total
Compton-thin AGN population (datapoints) compared with models m1
(dotted line) and m2 (dashed line). The soft XLF by HMS05 converted to
the hard band is also shown (solid line). The XLF datapoints by Ueda
et al. (2003) are limited to $z<3$.}
\label{ueda}
\end{figure*}

\begin{figure*}
\includegraphics[width=18cm]{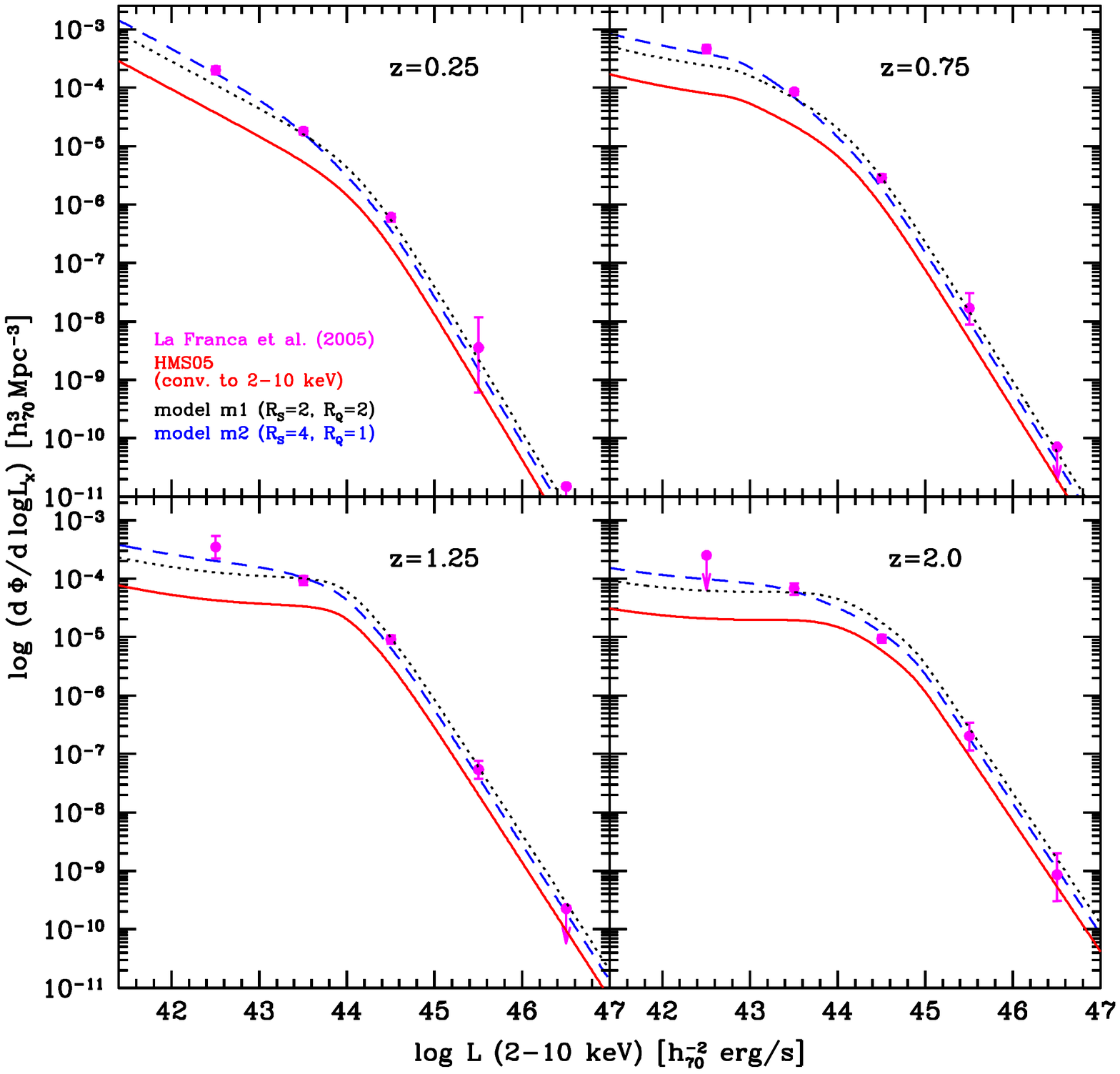}
\caption{Same as the previous Figure but considering the hard XLF data
of La Franca et al. (2005).}
\label{lafr}
\end{figure*}

\section{The XLFs and the obscured/unobscured AGN ratio}\label{xlfsect}

The accuracy in the determination of the AGN X-ray luminosity function
parameters has significantly improved in the past few years, leading
to a rather detailed knowledge of the AGN evolution up to redshifts of
about 4. Besides the soft XLF, whose first reliable determinations
date back to more than 10 years ago (Maccacaro et al. 1991; Boyle et
al. 1994, Page et al. 1996), it has been recently possible to obtain a
solid estimate of the AGN XLF also in the hard 2--10 keV band (Ueda et
al. 2003, La Franca et al. 2005, Barger et al. 2005). It is worth
noting that, although both the soft and hard X--ray selected sources
are well described by a LDDE model, the dependence of the evolution
rate with luminosity appears to be less extreme in the 2--10 keV
band. This can already be regarded as an indication of a variable
ratio between obscured and unobscured AGN as a function of luminosity
and/or redshift as it will be shown later. Also, combining several AGN
samples selected in the mid--infrared, optical blue band, soft and
hard X--rays, Hopkins, Richards \& Hernquist (2006) were able to
compute the evolution of a "bolometric" luminosity function (LF). A
luminosity dependent density evolution (LDDE) model, where the
evolution rate is higher for high-luminosity sources, combined with a
decreasing fraction of obscured sources for increasing luminosities,
provides a rather good description of AGN evolution over a wide range
of frequencies, bolometric luminosities and redshifts.
                                                                              
As a reference XLF for our modeling we chose the most up-to-date
calculation of Hasinger, Miyaji \& Schmidt (2005, hereafter HMS05) in
the 0.5--2 keV band, which is based on the largest number statistics
(about 1000 objects) and spectroscopic completeness. It contains
objects selected to have broad optical emission lines, or, when the
quality of the optical spectra was poor, on the basis of their soft
X-ray spectra (see HMS05 for details). It is known that about 10-20\%
of broad line AGN suffer from some X-ray absorption (e.g. Brusa et
al. 2003, Georgantopoulos et al. 2004), as well as that soft X-ray
spectra do not guarantee absence of absorption (especially for
high-redshift objects, for which the absorption cut off may be shifted
out of the observable X-ray band). Some contamination by obscured AGN
is then likely to affect the soft XLF, but this cannot be easily
sorted out and, at any rate, is not expected to alter significantly
the LDDE parameters. In the following we will then consider the HMS05
XLF as a good approximation to the soft XLF of X-ray unobscured
(log$N_H<21$) AGN.

While the soft XLF by HMS05 contains essentially unobscured AGN, the
hard XLFs by Ueda et al. (2003) and La Franca et al. (2005) do not
apply any pre-selection to the AGN included in their samples, which
therefore contain both populations of obscured and unobscured
AGN. Since both Ueda et al. (2003) and La Franca et al. (2005) measure
the column density for each object in their sample (via a rough X-ray
spectral analysis, often using hardness ratios), they can derive the
intrinsic, de-absorbed luminosity for each source as well as the
appropriate K-correction to the 2--10 keV rest frame band. They can
therefore determine the intrinsic, rest frame 2--10 keV XLF, which is
however dependent on the measured $N_H$ distributions, and may thus
suffer from selection biases towards less obscured objects. We
nonetheless consider the absorption corrections as a second order
effect on the integral XLF, which in the following will be considered
as independent on the measured $N_H$ distribution. The general
agreement between the XLF by Ueda et al. (2003) and La Franca et
al. (2003) despite the different $N_H$ distributions derived by these
authors supports this assumption.

\begin{figure}
\includegraphics[width=9cm]{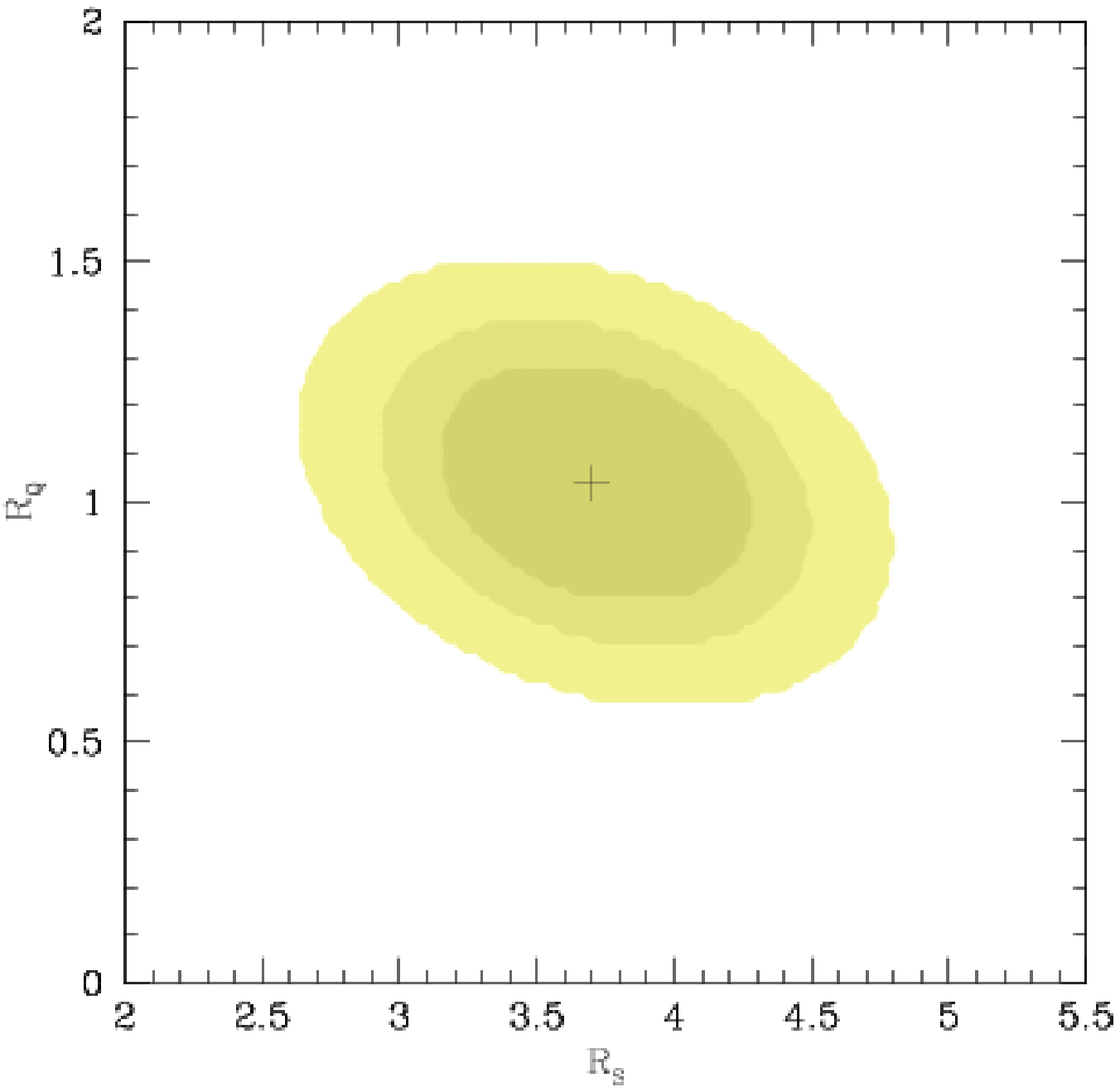}
\caption{The 68,90,99\% confidence contours on the best fit $R_S$ and
$R_Q$ ratios computed for two interesting parameters
($\Delta\chi^2=2.3,4.61,9.21$, respectively).}
\label{confcont}
\end{figure}

While the population of Compton-thin AGN should have been sampled with
a good accuracy level by the 2--10 keV surveys, heavily obscured
Compton-thick sources, which, almost by definition, are extremely
difficult to observe below 10 keV, should be essentially missing from
the hard XLFs. We will therefore consider the 2-10 keV XLF by Ueda and
La Franca as a good approximation to the 2-10 keV XLF of Compton-thin
(log$N_H<24$) AGN. As a consequence, the XLF of moderately obscured
AGN (21$<$log$N_H<$24) can be reasonably well estimated as the
difference between the hard XLFs and the soft XLF by HMS05, once the
latter is converted to the 2-10 keV band.


Since we assumed a distribution of spectra rather than a single
powerlaw, we have to take into account different band corrections. We
then split the soft XLF by HMS05 into 9 individual XLFs (as many
as the considered spectral powerlaws). Each of them was then
multiplied by the appropriate weight $p_i$ (see previous Section) and
converted to the 2--10 keV band with the corresponding spectral index.
The 2--10 keV converted individual XLFs were then summed together to
produce the total hard XLF for unobscured sources. As shown in
Fig.~\ref{ueda}, this is not sufficient to match the XLF data by Ueda
et al. (2003), and obscured AGN have to be added.

We introduce the ratio $R$ between obscured Compton-thin AGN and
unobscured AGN (i.e. the number ratio between sources with
$21<logN_H<24$ and with $logN_H<21$) defined as follows:

\begin{equation}
R(L) = R_S\;e^{-L/L_c} + R_Q\;(1-e^{-L/L_c}), \left\{
\begin{array}{ll}
R \rightarrow R_S \;\;\; (L<<L_c)\\
R \rightarrow R_Q \;\;\; (L>>L_c)\\
\end{array}
\right. ,
\label{ratio}
\end{equation}

where $R_S$ is the ratio in the Seyfert (low-) luminosity regime and
$R_Q$ is the ratio in the QSO (high-) luminosity regime, and $L_c$ is
the 0.5--2 keV characteristic luminosity dividing the two regimes (we
fixed $logL_c=43.5$). The total AGN hard XLF is simply obtained by
converting the soft (unobscured) XLF to the hard band and then
multiplying it by (1+$R(L)$).

We considered two hypotheses: first (model m1) a ratio independent of
luminosity (i.e. $R_Q=R_S$); second (model m2), a variable ratio,
where $R_S$ and $R_Q$ can vary independently.

We searched for the best fit $R_S$ and $R_Q$ values in model m1 and m2
by means of a $\chi^2$ test applied to the combined Ueda et al. (2003)
and La Franca et al. (2005) datapoints.\footnote{Although the data
samples used by Ueda et al. (2003) and La Franca et al. (2005) are not
completely independent from a statistical point of view, we note that
the output XLFs have been instead obtained with two completely
independent analysis methods. Therefore, we safely apply the $\chi^2$
test to the combined Ueda et al. (2003) and La Franca et al. (2005)
XLF datapoints. At any rate, when considering only the sample by La
Franca et al. (2005), which is the larger among the two, we obtained
$R_S$ and $R_Q$ values very similar to those from the combined fit.}
Model m1 is ruled out by the XLF comparison: the best fit ratio,
$R_S=R_Q=1.6$ ($\chi^2/dof=94.8/50=1.9$) is rejected at $>99.99\%$
confidence level. On the contrary a model with a varying ratio, with
best fit values $R_S=3.7$ and $R_Q=1.0$, provides a good match between
the soft and hard XLF and is statistically acceptable
($\chi^2/dof=51.1/49=1.0$), strongly supporting previous observations
of a declining ratio towards high intrinsic luminosities. We explored
in some more detail the issue of a variable ratio, in particular
trying to put some general constraints on the number of obscured
QSOs. We checked if a model without obscured QSOs at all is a viable
solution, by fixing $R_Q=0$ and leaving $R_S$ free to vary. The best
fit solution ($\chi^2/dof=103/50=2.1$) is clearly unacceptable. Also,
a large obscured QSOs fraction ($R_Q\geq 2$) does not provide an
acceptable solution. In Fig.~\ref{confcont} we show the 68,90 and 99\%
confidence contours on $R_S$ and $R_Q$ computed for two interesting
parameters ($\Delta\chi^2=2.3,4.61,9.21$, respectively; Avni 1976). In
the following we will consider the 99\% confidence intervals to quote
uncertainties on the $R_S$ and $R_Q$ values. These rather conservative
boundaries have been chosen to account for the uncertainties in the
LDDE parametrization of the input XLF by HMS05, which should be also
included in the error budget. The ratio $R_Q$ between obscured and
unobscured QSOs is then constrained to be within 0.6 and
1.5. Similarly $R_S$ is constrained to be in the range 2.6-4.8. As we
will show in the next Section a value of $R_Q\lesssim 1$ gives a
better agreement with the obscured QSO fractions observed in different
samples of 2-10 keV selected sources. In Fig.~\ref{ueda} and
Fig.~\ref{lafr} we show the XLF comparison with model m1 (where we
approximated $R_S=R_Q\sim 2$) and model m2 (where we approximated
$R_S\sim 4$). We also verified if the $R_S$ and/or $R_Q$ values could
increase with redshift, without finding any significant evidence. In
the following we will consider model m2 with $R_S=4$ and $R_Q=1$ as
our baseline model.

\begin{figure}[t]
\includegraphics[width=9cm]{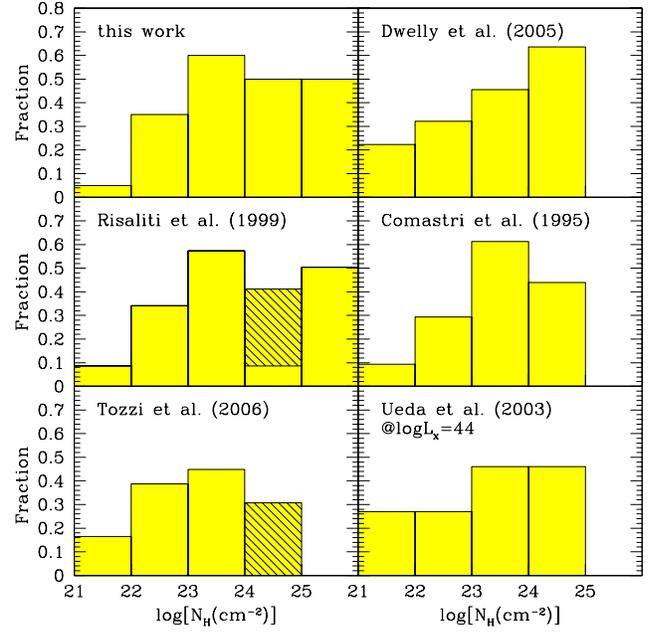}
\caption{Comparison between the model $N_H$ distribution (upper left)
and those obtained from previous works. The number of objects in each
$N_H$ bin is normalized to the total number of Compton-thin ($21<{\rm
log}N_H<24$) AGN, i.e. the sum of the first three $N_H$ bins is 1
(see Section~\ref{ncnd} for details). The estimated fractions of
Compton-thick objects are discussed in Sections 7 and 9. Shaded areas
refer to lower limits to $N_H$.}
\label{nhdist}
\end{figure}

\begin{figure}
\includegraphics[width=9cm]{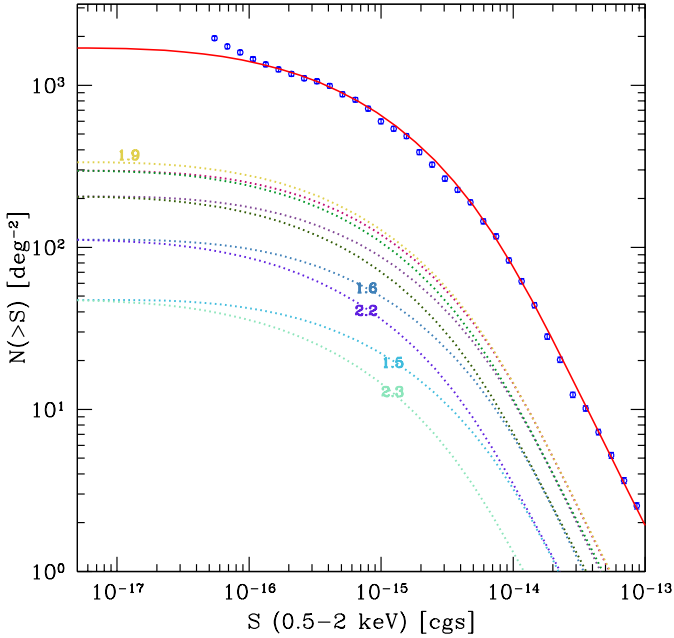}
\caption{Cumulative soft AGN counts for unabsorbed AGN. An input
distribution of photon indices with average=1.9 and dispersion=0.2 is
assumed. The individual logN-logS for the different spectral classes
(photon index is labeled for some of them), scaled by the corresponding
weight along the distribution, are shown as dotted lines. The total
logN-logS for unobscured AGN, resulting from the sum of the dotted
lines, is shown as a solid line and is compared with the soft
logN-logS measured for type-1 AGN by HMS05 (open circles).}
\label{sau}
\end{figure}

\begin{figure}
\includegraphics[width=9cm]{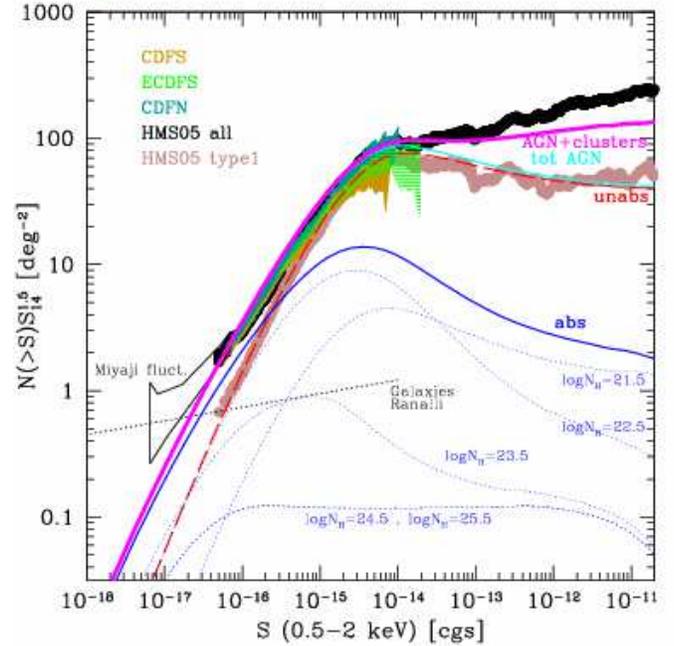}
\caption{Cumulative soft AGN counts normalized to an Euclidean
Universe. Datapoints are labeled on the top left: CDFS (Rosati et
al. 2002); E-CDFS (Lehmer et al. 2005); CDFN (Alexander et al. 2003);
compilation by Hasinger, Miyaji \& Schmidt (2005) for all the 0.5-2
keV X-ray sources (HMS05 all) and type-1 AGN only (HMS05 type-1). The
errorbox from the fluctuation analysis by Miyaji \& Griffiths (2002)
is also shown. The model curves for unabsorbed AGN, absorbed AGN,
total AGN, and total AGN plus clusters are shown respectively with the
following line styles: red long-dashed, blue solid, cyan solid,
magenta solid. The logN-logS curves for individual $N_H$ classes are
also shown as blue dotted lines. The curves for mildly (log$N_H=24.5$)
and heavily (log$N_H=25.5$) Compton-thick AGN are fully described in
Section~\ref{xrbsect}. The total data by HMS05 are not cleaned by
Galactic sources, thus producing an apparent discrepancy with respect
to the total model prediction at bright fluxes. The logN-logS for
normal galaxies by Ranalli et al. (2003) is also shown as a dotted
straight line, which is expected to provide a significant contribution
at very faint fluxes ($\lesssim 5 \times 10^{-17}$ cgs). }
\label{soft}
\end{figure}

\section{Number counts and $N_H$ distribution}\label{ncnd}

The comparison between the soft and hard XLF only constrains the total
number of obscured Compton-thin AGN, but does not provide informations
on their absorption distribution. The latter is estimated by
considering the soft and hard cumulative number counts as explained
below.

For each absorption class $j$ and slope $i$ individual logN-logS
relations $N_{ij}(>S)$ were computed integrating the XLF of HMS05
in the $10^{42}-10^{48}$ erg s$^{-1}$ 0.5-2 keV luminosity range and
0--5 redshift interval and taking into account the appropriate
K-corrections. Besides the K-correction another correction has to be
considered because of the shape of the effective area of X-ray imaging
instruments, which is maximum at 1--2 keV and therefore favors the
detection of sources with soft rather than hard spectral shapes. The
correction method, described in detail in Gilli, Salvati \& Hasinger
(2001), is briefly recalled here. We considered the effective area of
Chandra ACIS--I since many logN-logS, including the deepest ones, have
been obtained with this instrument, and computed the count rate
produced by sources with the same observed flux $S$ but different
absorptions and redshifts $CR(N_H,z)$. At any given flux $S$ the
logN-logS relation for each absorption class was computed at an
"effective" flux $S(N_H,z) = CR(\Gamma_{conv})/CR(N_H,z)$, where
$CR(\Gamma_{conv})$ is the count rate produced by a source with flux
$S$ and with the X-ray spectrum originally used to derive the
logN-logS measurement (i.e. to convert counts into fluxes). The net
effect is then such that the number counts of obscured sources were
computed at higher effective fluxes, thus mimicking the loss of
instrumental sensitivity towards hard spectra.

Each individual logN-logS for unobscured AGN was then scaled by a
factor $p_i$, where $p_i$ is the Gaussian weight for each 
spectral slope as defined in Eq.~\ref{pii}, while each individual
logN-logS for obscured AGN was scaled by a factor

\begin{equation}
f_{ij}=p_i\,R(L)\,d_j,
\label{wab}
\end{equation}

where $R(L)$ is the Compton-thin to unabsorbed AGN ratio defined in
Eq.~\ref{ratio} and $d_j$ is the fraction of obscured AGN in each of
the three Compton-thin classes log$N_H$=21.5,22.5,23.5, such that

\begin{equation}
\sum_{j=1}^3d_j=1
\end{equation}

Since a flat $N_H$ distribution was found to overestimate the
soft counts, we assumed that the factor $d_j$ is increasing with
obscuration from $logN_H=21.5$ to $logN_H=23.5$. A similar increase
has been actually observed by several authors using different
data-sets (e.g. Risaliti et al. 1999, Dwelly et al. 2005, Tozzi et
al. 2006) and already assumed in previous synthesis models (Comastri
et al. 1995, Gilli et al. 2001).

The considered $N_H$ distribution is shown in Fig.~\ref{nhdist},
together with other $N_H$ distributions from the literature. It is
worth noting that all of these distributions are intrinsic, i.e.
those which would be observed at extremely low (formally zero) limiting
fluxes.

With such an approach our model provides quantitative and accurate
predictions on both the $N_H$ and $\Gamma$ distribution to be observed
at any given limiting flux, thus allowing comparisons with the
numerous results obtained by recent X-ray surveys as discussed in
Section~\ref{add}.

\begin{figure}
\includegraphics[width=9cm]{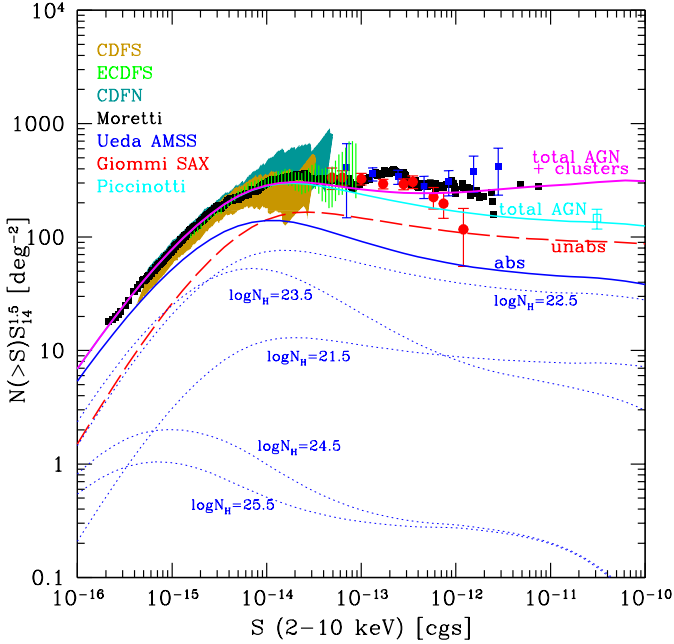}
\caption{Cumulative 2-10 keV AGN counts normalized to an Euclidean
Universe. The model curves and corresponding line styles are the same
as in Fig.~\ref{soft} and are compared with datapoints from different
surveys: CDFS (Rosati et al. 2002); E-CDFS (Lehmer et al. 2005); CDFN
(Alexander et al. 2003); the Moretti et al. (2004) compilation; ASCA
Medium Sensitivity Survey (AMSS, Ueda et al. 2001); BeppoSAX 2-10 keV
survey (Giommi et al. 2002); HEAO-1 A2 all sky survey (Piccinotti et
al. 1982). The logN-logS curves for mildly (log$N_H=24.5$) and heavily
(log$N_H=25.5$) Compton-thick AGN are fully described in
Section~\ref{xrbsect}.}
\label{hard}
\end{figure}

\begin{figure}
\includegraphics[width=9cm]{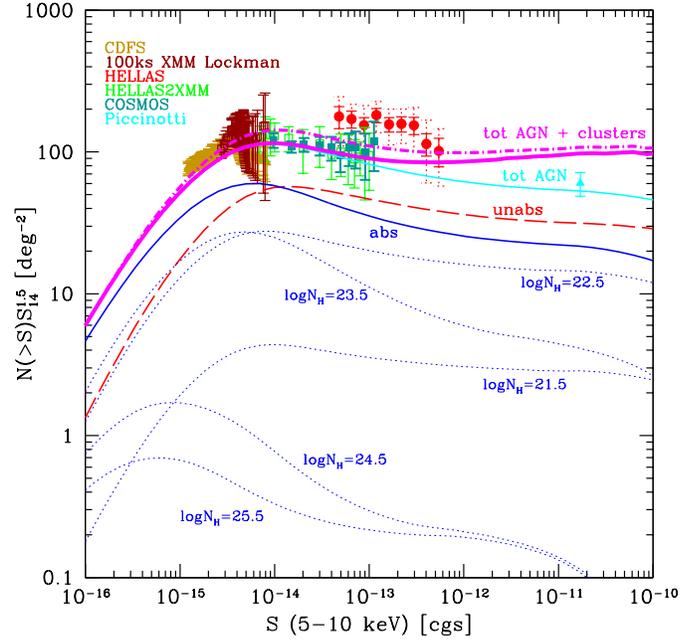}
\caption{Cumulative 5-10 keV AGN counts normalized to an Euclidean
Universe. The model curves and corresponding line styles are the same
as in the previous figures and are compared with datapoints from
different surveys: CDFS (Rosati et al. 2002); 100 ks Lockman Hole
(Hasinger et al. 2001); BeppoSAX HELLAS (Fiore et al. 2003);
HELLAS2XMM (Baldi et al. 2002); COSMOS (Cappelluti et al. 2006);
HEAO-1 A2 all sky survey (Piccinotti et al. 1982) converted to the
5-10 keV band. The logN-logS curves for mildly (log$N_H=24.5$) and
heavily (log$N_H=25.5$) Compton-thick AGN are fully described in
Section~\ref{xrbsect}. The dot-dashed line is the total model
prediction when assuming an average spectral index of
$\langle\Gamma\rangle=1.8$ (see the Discussion).}
\label{vhard}
\end{figure}

As shown in Fig.~\ref{nhdist} the $N_H$ distribution considered in
this work contains a very low fraction of AGN obscured by column
densities in the range $21<logN_H<22$. While this may appear
unrealistic, it should be noted that about 10\% to 20\% of the sources
in the HMS05 type-1 AGN catalog, which here are assumed to be
unobscured, may belong to the class of mildly obscured 21$<log N_H<$22
AGN. Absorbing column densities in this range may well have escaped
detection especially if at moderate to high redshifts. These AGN would
mostly ``fill in'' the $logN_H=21.5$ bin, thus making the steep rising
towards high columns less extreme than that shown in
Fig.~\ref{nhdist}. If this is the case, the ratio between obscured and
unobscured AGN determined in the previous Section would increase by
$\sim 20-30\%$.

In Fig.~\ref{sau} we show the individual 0.5-2 keV logN-logS for the
unobscured AGN population, each scaled by the appropriate factor
$p_{i}$ as a function of different spectral slope, where the effects
of the different K-corrections for the different slopes can be fully
appreciated. As an example, sources with $\Gamma=1.5$ and
$\Gamma=2.3$, which are weighted by the same factor $p_i$ for a
Gaussian distribution centered at $\langle\Gamma\rangle$=1.9, provide
an equal contribution at very faint fluxes (as it should be expected
since at very faint fluxes all the sources in the XLF can be
detected), while at bright fluxes objects with $\Gamma=1.5$ are
observed more easily due to their stronger K--correction.

The 0.5--2 keV AGN counts resulting from our baseline model m2 are
shown in Fig.~\ref{soft}, where we plot the logN-logS relation scaled
by $S_{14}^{1.5}$ (where $S_{14}$ is the flux in units of $10^{-14}$
erg cm$^{-2}$ s$^{-1}$) to expand the scale on the y-axis and
facilitate the evaluation of any difference between the data and the
model. The source counts for unobscured and Compton-thin AGN as well
as the total are shown. In addition, we also computed the contribution
from galaxy clusters following Gilli et al. (1999b). From an
observational point of view, the logN-logS relation in the soft band
is now measured over a flux range encompassing more than 6 orders of
magnitude (HMS05). A compilation from surveys at different limiting
fluxes is shown in Fig.~\ref{soft}. The model logN-logS is found to
reproduce with excellent accuracy the source counts over the whole
flux range with the exception of very bright and very faint
fluxes. This has however to be expected since X--ray sources other
than AGN contribute in those regimes: at bright fluxes ($S>10^{-13}$
cgs) the total logN-logS by HMS05 includes also galactic sources
(therefore the total model logN-logS -- AGN plus clusters --
underestimates the observed counts); at very faint fluxes ($\lesssim
5\,10^{-17}$ cgs) a population of normal and star forming galaxies is
expected to contribute significantly to the total counts (see
e.g. Ranalli et al. 2003). The galaxy logN-logS predicted by Ranalli
et al. (2003) is also plotted in Fig.~\ref{soft}


\begin{figure}[t]
\includegraphics[width=9cm]{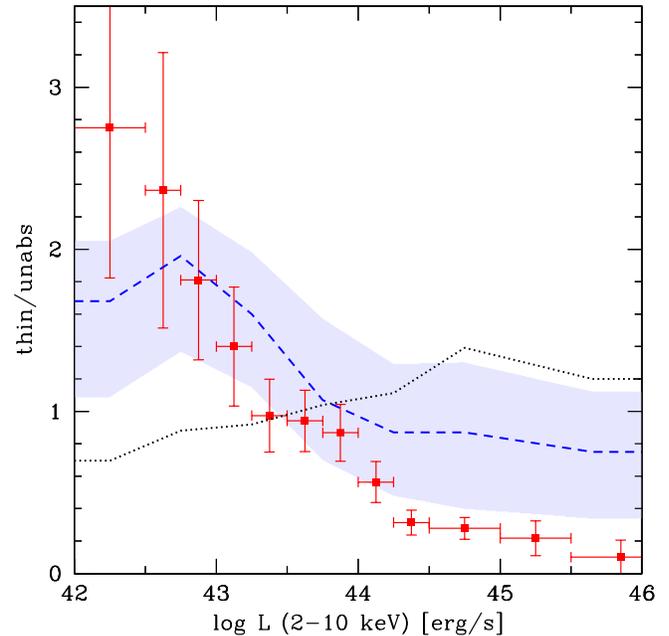}
\caption{The observed ratio between optical type-2 and type-1 AGN in
the Hasinger (2006) sample as a function of intrinsic 2-10 keV
luminosity compared with ratio between obscured (log$N_H>21$) and
unobscured (log$N_H<21$) AGN as predicted by model m2 (dashed line)
and m1 (dotted line) after folding with the appropriate selection
effects of the observed sample. The shaded area represent the
uncertainties on model m2. Model m1 (constant intrinsic ratio)
appears strongly disfavored.}
\label{grhrat}
\end{figure}

\begin{figure}[t]
\includegraphics[width=9cm]{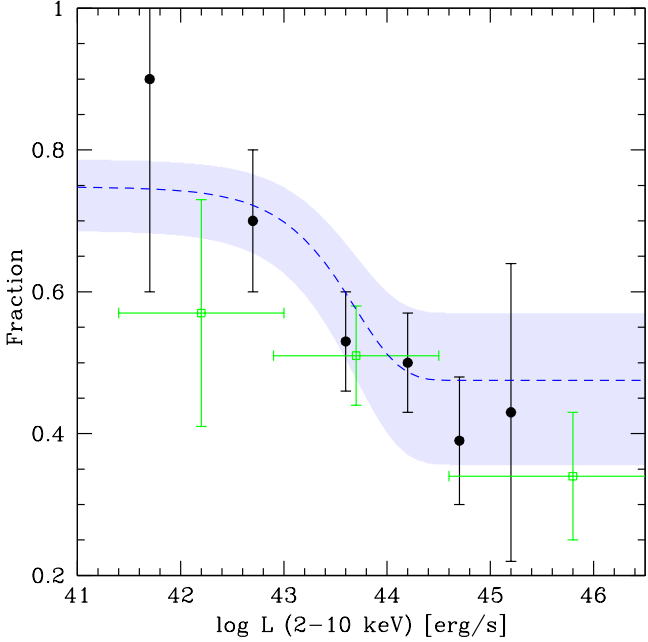}
\caption{The intrinsic fraction of AGN with log$N_H>22$ vs intrinsic
2-10 keV luminosity as determined by Akylas et al. (2006, filled
circles) and Ueda et al. (2003, open squares) compared with the
corresponding fraction assumed in model m2 (dashed
line). Uncertainties on model m2 are shown by the shaded area.}
\label{akyrat}
\end{figure}

We then checked the model expectations against the source counts in
the harder 2--10 keV and 5--10 keV bands. The results are shown in
Fig~\ref{hard} and Fig~\ref{vhard}, respectively, where we also plot
the individual curves for absorbed and unabsorbed AGN, total AGN and
total AGN plus galaxy clusters. At fluxes above $\sim 10^{-13}$ cgs
galaxy clusters provide a significant contribution both in the 2-10
keV and 5-10 keV bands. The datapoint at $\sim 3\,10^{-11}$ cgs
represents the surface density of AGN in the Piccinotti et al. (1982)
sample, which is well matched by the model curve representing AGN
alone. While both the 0.5--2 keV and the 2--10 keV source counts are
well reproduced over several orders of magnitude in flux, in the 5--10
keV band the model predictions appear in good agreement with the
observations in the flux range covered by the HELLAS2XMM (Baldi et
al. 2002) and COSMOS (Cappelluti et al. 2006) surveys, but, on
average, lay slightly below the available measurements (although
within 1-2$\sigma$). The very hard 5--10 keV counts are almost
perfectly fitted assuming a slightly harder average spectrum
($\langle\Gamma\rangle=1.8$; see Fig.~\ref{vhard}). However, it is
likely that non-negligible systematic uncertainties affect the
logN--logS of sources detected in the narrow 5--10 keV band where the
instrumental sensitivity rapidly falls off. We will return on this
point in the Discussion.

Once the total number and $N_H$ distribution of the Compton-thin AGN
population have been fixed, we can now predict the fraction of
obscured AGN at different fluxes and in different luminosity bins and
compare it with the observations. In particular we considered the
compilation of 2--10 keV selected AGN by Hasinger (2006 in
preparation; see also Hasinger 2004), who combined the data from
surveys of different areas and limiting fluxes to get a sample of
$\sim 700$ objects. In Fig.\ref{grhrat} we show the ratio between
optical type-2 and type-1 AGN as a function of their intrinsic 2--10
keV luminosity as obtained by Hasinger (2006), compared with the
expectations of models m1 and m2 folded with an appropriate
sensitivity curve to account for the selection effects in the observed
sample. In model m1, where the ratio between obscured and unobscured
sources is constant, the $observed$ ratio appears to increase towards
higher luminosities, since obscured sources of low intrinsic
luminosity are the first to be missed in flux limited samples. In
model m2 the observed trend is reversed with respect to model m1
because the above mentioned effect is compensated by the intrinsically
higher ratio at low luminosities. As it is evident, model m1 is in
stark contrast with the observations, while model m2 produces a much
better agreement, although the model predictions are on average
slightly higher than the data. We note that several uncertainties
affect the comparison shown in Fig.~\ref{grhrat}: first of all the
sensitivity curve adopted to get the model predictions is just an
approximation to the full sky coverage of the Hasinger (2006) sample;
then comparing an optical type-2/type-1 ratio with an X-ray obscured to
unobscured AGN ratio always suffers from limitations related to the
AGN classification in different bands; finally a $\sim 20\%$ degree of
spectroscopic incompleteness is present in the considered
sample. Therefore, we do not try to accurately reproduce the observed
fractions but we use this plot to further rule out model m1.

The decreasing obscured AGN fraction towards high luminosities assumed
in model m2 is also in good agreement with the very recent
determinations of Akylas et al. (2006) who combined a number of
shallow XMM pointings with the Chandra Deep Field South data to derive
the intrinsic fraction of obscured AGN as a function of their
intrinsic 2--10 keV luminosity. As shown in Fig.\ref{akyrat}, the
fraction of AGN with log$N_H<22$ found by Akylas et al. (2006) is well
matched by our baseline model m2 (the same fraction as measured by
Ueda et al. 2003 is also reported in the plot.)

\begin{figure}[t]
\includegraphics[width=9cm]{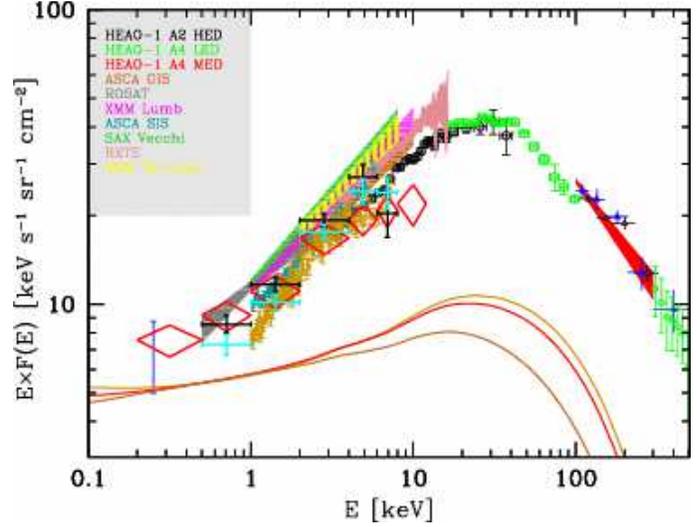}
\caption{The integrated contribution of unobscured AGN to the XRB
spectrum as a function of the dispersion in their spectral
distribution ($\sigma_{\Gamma}$). From bottom to top curve:
$\sigma_{\Gamma}$ = 0.0, 0.2, 0.3. The spectral distribution is
centered at $\langle\Gamma\rangle=1.9$. Datapoints are explained in
detail in the Caption of Fig.~\ref{xrb}.}
\label{xrbdisp}
\end{figure}

\begin{figure*}
\includegraphics[width=9cm]{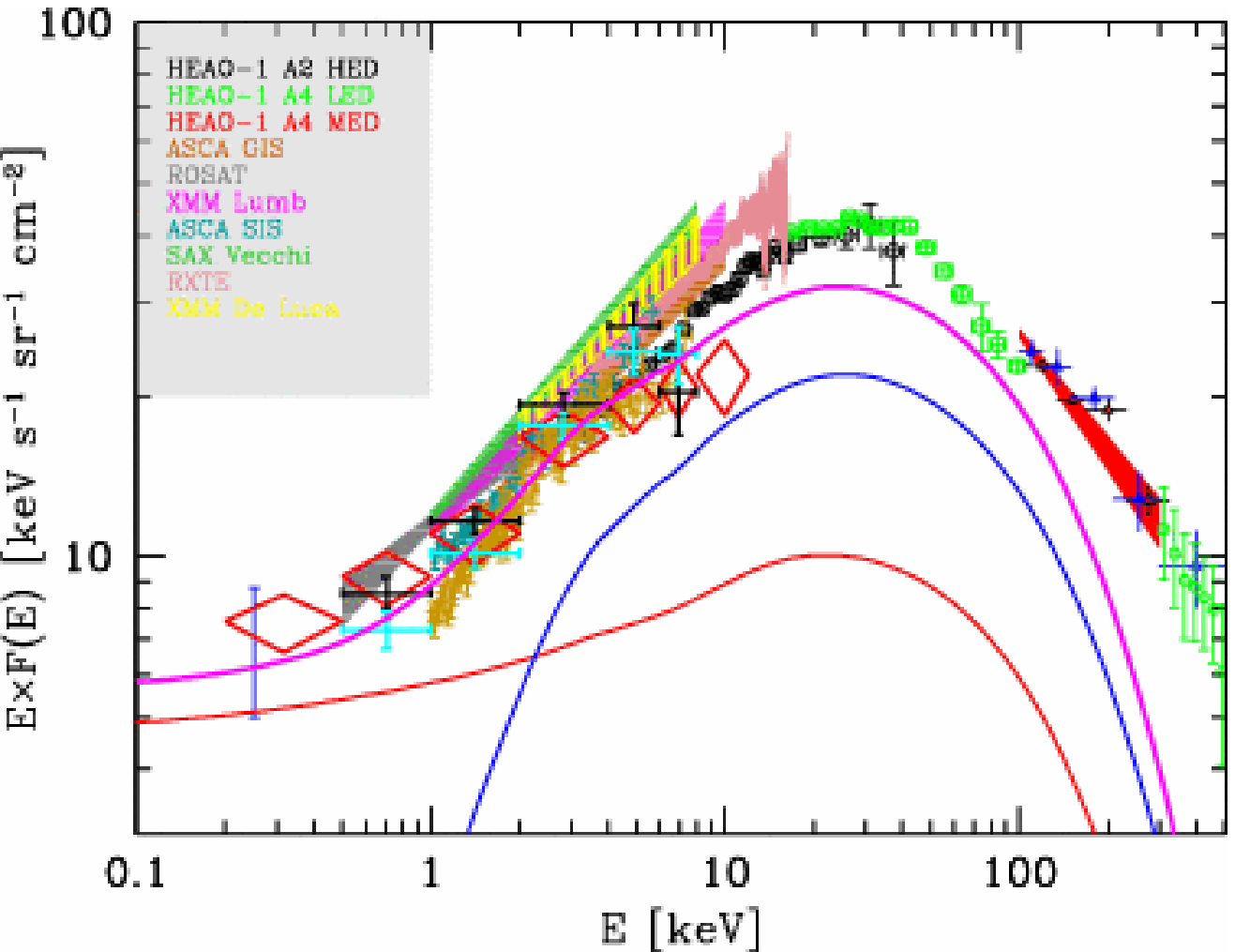}
\includegraphics[width=9cm]{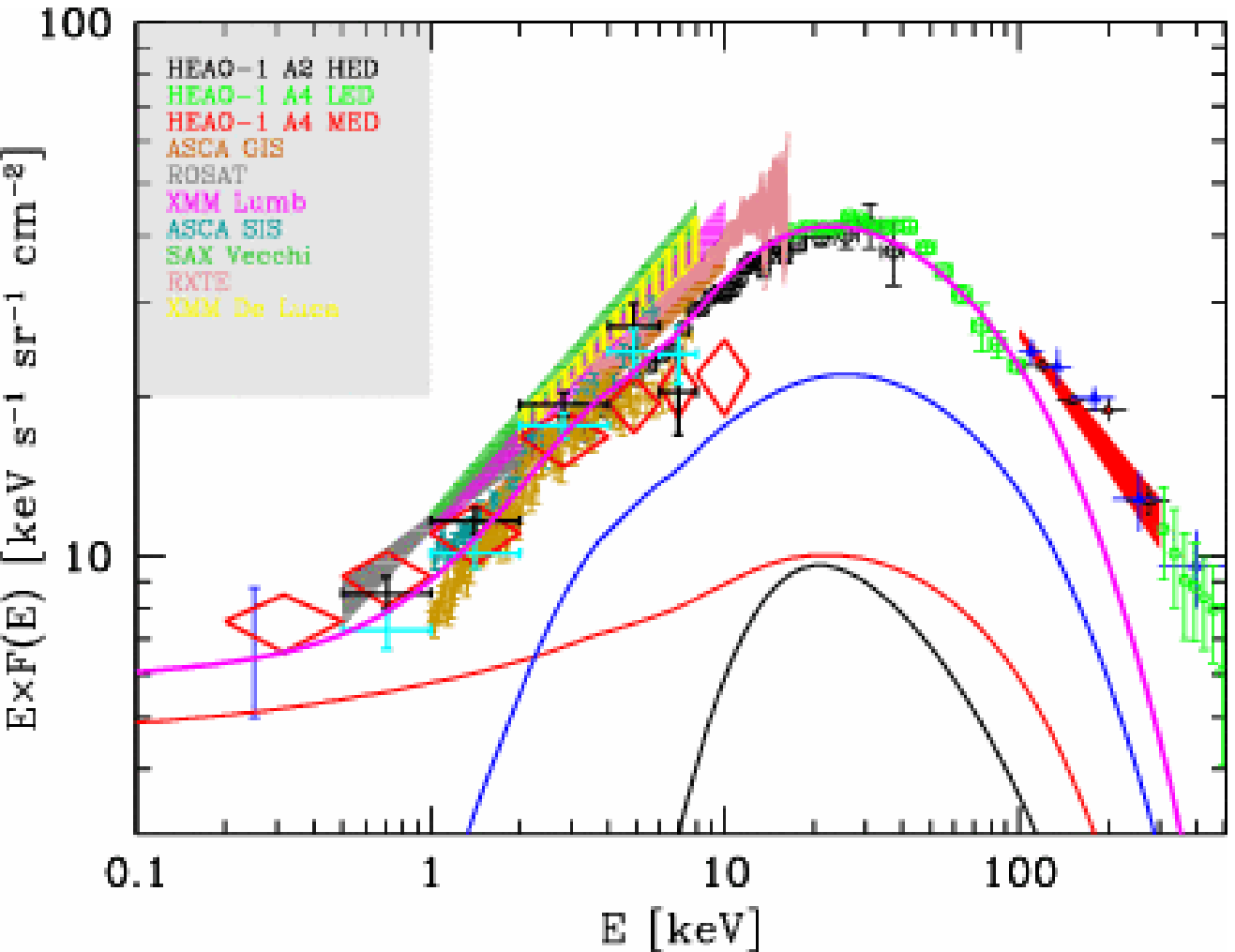}
\caption{(a): The cosmic XRB spectrum and predicted contribution from
the population of Compton-thin AGN. The different XRB measurements are
explained on the top left: different instruments on board HEAO-1
(Gruber 1992, Gruber et al. 1999); ASCA GIS (Kushino et al. 2002);
ROSAT PSPC (Georgantopoulos et al. 1996); two different measurements
by XMM (Lumb et al. 2002, De Luca \& Molendi 2003); ASCA SIS (Gendreau
et al. 1995); BeppoSAX (Vecchi et al. 1999); RXTE (Revnivtsev et
al. 2003). At $E>100$ keV the plotted datapoints are from HEAO-1 A4
MED (red triangles: Gruber 1992, Gruber et al. 1999; shaded area:
Kinzer et al. 1997); balloon experiments (blue triangles, Fukada et
al. 1975); SMM (green circles, Watanabe et al. 1997). The blue
errorbar at 0.25 keV is from shadowing experiments by Warwick \&
Roberts (1998). Also shown are the XRB fractions resolved by Worsley
et al. (2005) in the Lockman Hole (red diamonds), CDFS (cyan crosses)
and CDFN (black crosses). The resolved fraction in the CDFS as
measured by Tozzi et al. (2001a) is also shown (gold
datapoints). Solid lines refer to the contribution of different AGN
classes according to model m2. Unobscured AGN, obscured Compton-thin
AGN, total AGN plus galaxy cluster are shown with a red, blue and
magenta curve, respectively. (b): same as the previous panel but
including also the contribution of Compton-thick AGN (black line).}
\label{xrb}
\end{figure*}

\section{The XRB spectrum}\label{xrbsect}

Having obtained a robust description of the unobscured and
Compton-thin AGN properties -- in particular of their relative ratio
and absorption distribution -- making use of the soft and hard XLF and
logN--logS, their contribution to the XRB spectrum can be readily
computed.  As in the case of the logN-logS relations, we computed the
contribution from each individual spectral and absorption class using
the appropriate weight as given by Eq.~\ref{wab}, and then summed all
of them together. Consistently with the computation of the source
counts (see Section~\ref{ncnd}), the input XLF of HMS05 is integrated
in the $10^{42}-10^{48}$ erg s$^{-1}$ 0.5-2 keV luminosity range and
0--5 redshift interval.

In order to quantify the effects of an intrinsic dispersion in the
spectral slope distribution we show in Fig.~\ref{xrbdisp} the
contribution of only unobscured AGN. Increasing the dispersion from
zero to 0.2 and 0.3, the contribution at 30--40 keV is increased by
30--40\%. This has the obvious consequence of reducing the number of
obscured sources required to match the XRB peak intensity.

The effects of the dispersion in the photon indices is considered also
when computing the contribution of the absorbed Compton-thin
population. In Fig.~\ref{xrb}a we show the separate contribution to
the XRB spectrum from unobscured and obscured Compton-thin AGN as well
as their summed contribution. The total curve also includes the
contribution from galaxy clusters (which is at most $\sim 10\%$ at 1
keV; see Gilli et al. 1999b).

The integrated emission from the considered populations reproduces the
entire resolved XRB flux (e.g. Worsley et al. 2005) up to 5--6 keV,
while in the 6--10 keV range it is slightly above it. In other words,
the deepest X--ray surveys have already sampled the whole unobscured
and most of the obscured Compton-thin population.  Also, Compton-thin
AGN are found to explain most of the XRB below 10 keV as measured by
HEAO-1 A2, but fail to reproduce the 30 keV bump, calling for an
additional population of Compton-thick objects. We then added as many
Compton-thick AGN as required to match the XRB intensity above 30 keV,
under the assumptions that the number of mildly Compton-thick objects
(log$N_H$=24.5) is equal to that of heavily Compton-thick objects
(log$N_H$=25.5), as observed in the local Universe (Risaliti et
al. 1999), and that they have the same cosmological evolution of
Compton-thin AGN. The fit requires a population of Compton-thick AGN
as numerous as that of Compton-thin ones, i.e. four times the number
of unobscured AGN at low luminosities and an equal number at high
luminosities. With Compton-thick AGN the total obscured to unobscured
AGN ratio decreases from 8 at low luminosities to 2 at high
luminosities.  It is worth noting that the above ratio is estimated by
fitting the XRB level at 30 keV as measured by HEAO1--A2. The global
fit to the XRB spectrum is shown in Fig.~\ref{xrb}b. Since the HEAO-1
A2 background is found to be lower by about 20\% with respect the most
recent determinations of the 2--10 keV background intensity (Kushino
et al. 2002, Lumb et al. 2002, Hickox \& Markevitch, 2006), our model
does not account for the full XRB values measured by e.g. ASCA, XMM
and Chandra. We will address the issue of the XRB spectral intensity
in the Discussion.

Having constrained the space density of Compton-thick AGN with the fit
to the XRB, the source counts in the 0.5--2 keV, 2--10 keV and 5--10
keV can be computed for the entire AGN population. Although
Compton-thick AGN provide a measurable contribution only at very faint
fluxes (see Figs~\ref{soft},\ref{hard},\ref{vhard}), it is interesting
to look at the behaviour of their logN-logS in more detail. In the
soft band (see Fig.\ref{soft}) the curves for mildly and heavily
Compton-thick AGN coincide since i) their space density is the same
and ii) they have the same K-correction. Indeed, since the spectrum of
mildly and heavily Compton-thick AGN is the same (reflection
dominated) up to $\sim 10$ keV (see Fig.\ref{xspec}), the 0.5-2 keV
band is sampling an identical continuum even for sources at high
redshift (up to $z\sim 4$). In the 2--10 keV and 5--10 keV band
instead the curves for mildly Compton-thick and heavily Compton-thick
sources show significant differences: at very bright fluxes, above
$\sim 10^{-12}$ cgs, where only local sources are visible, the
logN-logS curves of the two Compton-thick classes coincide because in
the 2--10 keV rest frame band their spectrum is dominated by the same
reflection continuum (Fig.~\ref{xspec}). On the contrary, at fainter
fluxes, $\sim 10^{-14}-10^{-15}$ cgs, where more distant sources can
be detected, the surface density of mildly Compton-thick AGN appears
about twice that of heavily Compton-thick AGN because of the stronger
K-correction produced by the transmitted continuum (Fig~.\ref{xspec}).




\section{Additional constraints}\label{add}

\subsection{The observed fractions of obscured and Compton-thick AGN}

There is strong evidence, obtained combining deep and shallow surveys
over a broad range of fluxes, of an increasing fraction of obscured
AGN towards faint fluxes (see e.g. Piconcelli et al. 2003). This
general trend was expected and predicted by AGN synthesis
models. However, the very steep increase in the observed ratio from
bright to faint fluxes is poorly reproduced by models where the
obscured to unobscured AGN ratio does not depend on X--ray luminosity
(see Comastri 2004 for a review), while it is best fitted by assuming
that the obscured AGN fraction increases towards low luminosity and/or
high redshifts (La Franca et al. 2005).

\begin{figure}[t]
\includegraphics[width=9cm]{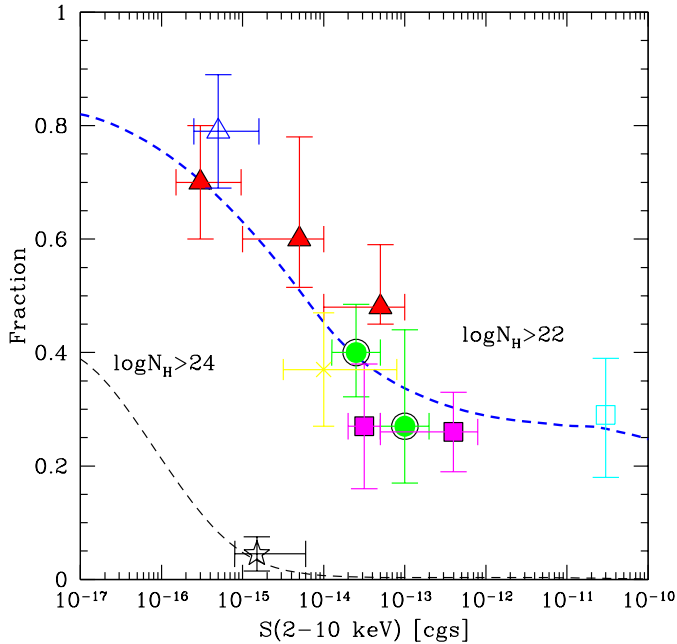}
\caption{The fraction of obscured AGN observed in different X-ray
surveys as a function of 2-10 keV limiting flux compared with the
model predictions. The upper (lower) curve and datapoints refer to the
fraction of obscured AGN with log$N_H>22$ (log$N_H>24$). Datapoints
have been collected from the CDFS (open triangle and star, Tozzi et
al. 2006), CDFN (filled triangles, Barger et al. 2005), XMM-Lockman
Hole (cross, Mainieri et al. 2002), HELLAS2XMM (filled circles, Perola
et al. 2004), Piconcelli et al. (2003) sample (filled squares),
Piccinotti et al. (1982) sample (open square). The $N_H$ measurements
for the Piccinotti et al. (1982) sample have been drawn from Shinozaki
et al. (2006).}
\label{picoplot}
\end{figure}

We compare the observed fraction of AGN with log$N_H>22$ with the
model predictions in Fig.~\ref{picoplot}. The choice of an absorption
threshold at log$N_H>22$ rather than at log$N_H>21$ provides a more
solid observational constraint, given the uncertainties in revealing
mild absorption in sources at moderate to high redshift and/or with
low photon statistics (Tozzi et al. 2006, Dwelly et al. 2005). The
model curve is able to reproduce the steep increase of the absorbed
AGN fraction from about 20--30\% at $\lesssim 10^{-13}$ cgs, i.e. at
the flux level of ASCA and BeppoSAX medium sensitivity surveys, to
70-80\% as observed at $5\,10^{-15}$ cgs in the deep Chandra
fields. Recently, Tozzi et al. (2006) performed a detailed X-ray
spectral analysis of the CDFS sources, identifying 14 objects, i.e.
about 5\% of the sample, as likely Compton-thick candidates. As shown
in Fig.\ref{picoplot}, this measurement is found to be in excellent
agreement with the fraction of Compton-thick AGN predicted by our
model at that limiting flux. These results confirm that below 10 keV
the large population of Compton-thick sources is poorly sampled even
by the deepest surveys.

\begin{figure}[t]
\includegraphics[width=9cm]{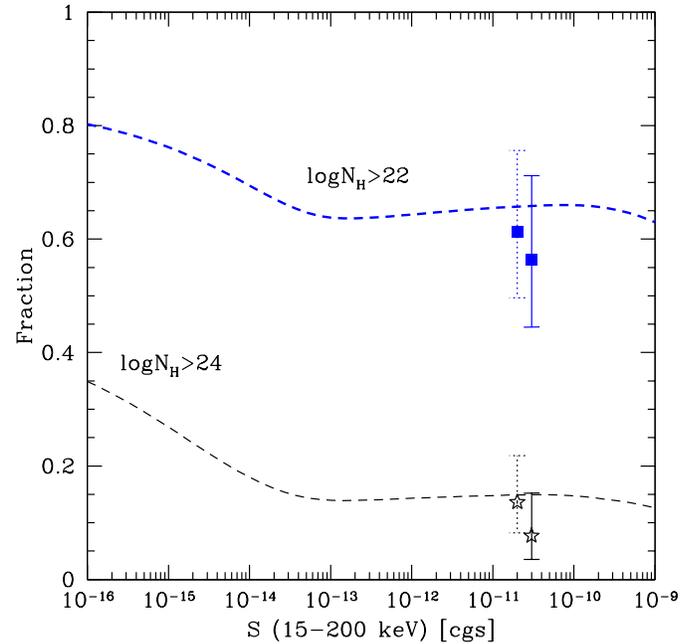}
\caption{The fraction of obscured sources as a function of the 15-200
keV limiting flux predicted by the baseline model m2. The upper and
lower curves refer to the fraction of objects with log$N_H>22$ and
log$N_H>24$, respectively. Datapoints show the corresponding fractions
found in the first Swift/BAT AGN catalog. While the datapoints with
solid errorbars show the actually measured fraction, datapoints with
dotted errorbars assume that most of the unidentified sources are
obscured.}
\label{bat}
\end{figure}

Very recently the first statistically well defined samples of AGN
selected at energies above 10 keV have become available. The first
release of AGN catalogs detected by the IBIS (20--100 keV band) and
ISGRI (20--40 keV band) instruments on board INTEGRAL (Bird et
al. 2006; Beckmann et al. 2006) includes about 40-60 objects. At the
bright fluxes sampled by INTEGRAL (a few times $10^{-11}$ cgs in the
20-40 keV band), about two thirds of the identified AGN are absorbed
by a column density in excess of log$N_H>22$ and about 10-15\% have
been found to be Compton-thick (Beckmann et al. 2006; Bassani et
al. 2006). While the quoted fractions should be taken with the due
care since the INTEGRAL AGN samples are still incomplete, they
nonetheless are in good agreement with those measured at similar
fluxes and for a similar waveband (15--200 keV) in the first Swift/BAT
AGN catalog (Markwardt et al. 2006), which on the contrary is about
90\% complete.

By considering the X-ray properties of the unidentified
sources, the number of Compton-thick AGN in the first Swift/BAT catalog
is estimated to be between 3 and 6, which translates into a fraction
of 7-14\%.


We computed the logN-logS relations expected from our baseline model
m2 in the Swift/BAT band, as well as the expected fractions of
obscured sources. At a limiting flux of $3\,10^{-11}$ cgs in the
15-200 keV band, i.e. the limiting flux of the BAT AGN catalog, about
65\% of the AGN are obscured by log$N_H>22$ and about 15\% are
Compton-thick, in excellent agreement with the observations (see
Fig.~\ref{bat}).

%

\subsection{Spectral distribution}

\begin{figure}[t]
\includegraphics[width=9cm]{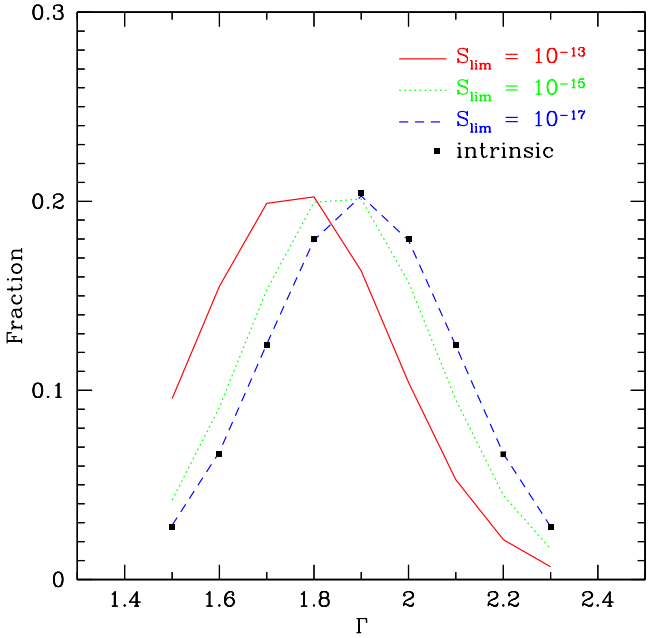}
\caption{The spectral distribution of unobscured AGN to be observed at
different 2-10 keV limiting fluxes as predicted by model m2. At very
faint fluxes ($\sim 10^{-17}$ cgs) all the sources can be detected and
the observed distribution coincides with the intrinsic one (filled
squares) with $\langle\Gamma\rangle=1.9$ and $\sigma_{\Gamma}=0.2$.}
\label{dispflux}
\end{figure}

An interesting feature of having assumed a spectral index distribution
is that the expected average spectral index is a function of the
survey limiting flux. This can be easily understood by looking at
Fig.~\ref{sau}, where the logN-logS for unobscured AGN with different
spectral slope is plotted. Due to the different K-corrections, at
bright X-ray fluxes sources with harder spectrum are detected more
easily, while at fainter fluxes, where most of the sources in the XLF
are being sampled, the observed distribution approaches the assumed,
intrinsic one. In Fig.~\ref{dispflux} we show the spectral
distribution expected at different 2-10 keV fluxes for unobscured
AGN. While at fluxes below $10^{-13}$ cgs the average expected slope is
$\Gamma=1.7-1.8$, the spectral distribution progressively moves
towards steeper values at fainter fluxes, until it exactly overlaps
with the assumed input one at very faint fluxes $\sim 10^{-17}$ cgs.

\begin{figure}[t]
\begin{center}
\includegraphics[width=7cm]{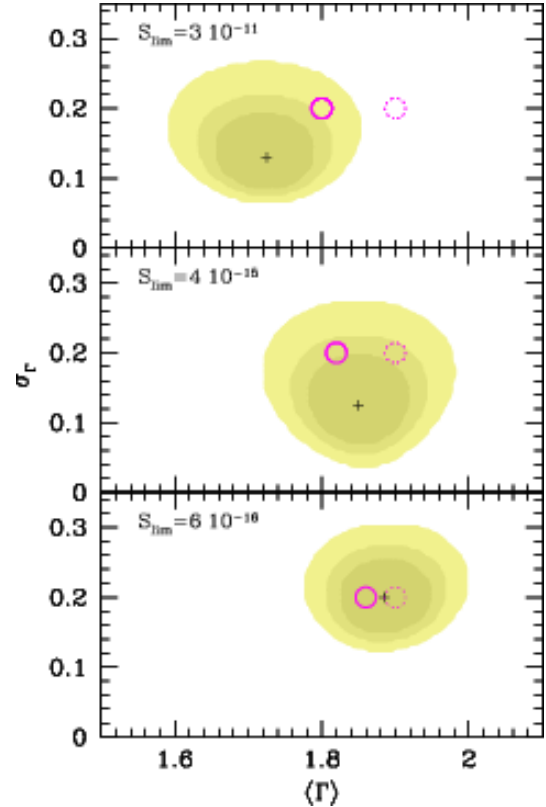}
\caption{The average photon index and intrinsic dispersion measured in
different samples of type-1 AGN (shaded areas; from top to bottom:
Shinozaki et al. 2006, Tozzi et al. 2006, Mateos et al. 2005),
compared with the model predictions for unobscured AGN (solid circles)
at the corresponding 2-10 keV limiting fluxes (upper left
labels). Only sources with good photon statistics have been
considered. The different contours refer to the 68,90 and 99\%
confidence level for two interesting parameters
($\Delta\chi^2=2.3,4.61,9.21$, respectively) around the best fit
values (crosses). Contours in the lower panel have been adapted from
Mateos et al. (2005). The dotted circles represent the intrinsic
(i.e. at zero flux) average photon index and dispersion assumed in the
model.}
\label{dispmateos}
\end{center}
\end{figure}

This expected trend can be checked against the properties of sources
detected in surveys at different limiting fluxes, provided that the
photon statistics in the observed X-ray spectra is sufficiently high
to provide a good estimate of the spectral slope. The X-ray spectral
sample presented by Mateos et al. (2005) for AGN detected at 2-10 keV
fluxes above $\sim 6\,10^{-16}$ cgs in the 800 ksec XMM-Lockman Hole
satisfies this requirement since all AGN in their sample have been
selected to have at least 500 X-ray counts in the 0.2-12 keV band. To
reduce at minimum the uncertainties on the spectral slopes which, for
sources with the same photon statistics, are larger for absorbed X-ray
spectra due to the correlation between absorption and powerlaw index,
we considered only the 46 type-1 AGN in Mateos et al. (2005), for
which they measure an average spectral slope of $\Gamma=1.89\pm0.06$
with an intrinsic dispersion of $\sigma_{\Gamma}=0.20\pm0.04$ (see
their Table~4) using the maximum likelihood method described by
Maccacaro et al. (1988). We also considered two additional source
samples at different fluxes. First we considered the CDFS sample by
Tozzi et al. (2006), by selecting only spectroscopically confirmed
type-1 AGN with more 500 X-ray counts. A total of 22 sources were
selected in this way, with 2-10 keV fluxes above $4\,10^{-15}$
cgs. Then, at much brighter fluxes we considered the sample presented
by Shinozaki et al. (2006) who analyzed the X-ray spectra, mainly
obtained with XMM, of the Piccinotti et al. (1982) sample. This
includes 17 type-1 AGN detected by HEAO-1 A2 at 2-10 keV fluxes above
$3\,10^{-11}$ cgs, for which even short XMM exposures provide good
quality data and allow a careful spectral analysis. For both samples
we computed the average photon index and dispersion using the same
Maximum Likelihood method described by Maccacaro et al. (1988). The
results, combined with those obtained by Mateos et al. (2005) and
compared with the model predictions for unobscured AGN at the same
limiting fluxes, are shown in Fig.~\ref{dispmateos}. The average
steepening of the unobscured AGN slope towards faint X-ray fluxes
predicted by the model appears to be in good agreement with the
observations.  It is worth stressing that the above described
steepening of the intrinsic source spectra towards faint fluxes is due
to the assumption of an intrinsic dispersion of the average power law
slopes and is by no means in contrast with the flattening of the
average slope observed in deep fields (Tozzi et al. 2001b). The latter
is the result of fitting with a single power law the spectra of
obscured sources and is ruled by their increasing contribution towards
faint fluxes.

\section{Discussion}

The results of the self-consistent XRB modelling presented in the 
previous sections have been obtained under several, though observationally 
justified, assumptions. In the following we will further explore the parameter 
space in order to strengthen the model predictive power especially 
for what concerns the population of Compton-thick AGN.

\subsection{Exploring the parameter space}\label{explo}

The choice of a distribution of AGN X--ray spectra with an average
slope $\langle\Gamma\rangle=1.9$ and dispersion $\sigma_{\Gamma}$=0.2,
though consistent with the observations of bright AGN, may not
necessarily hold over the broad range of redshifts and luminosities
here considered. In the following it will be shown that, within
the basic model assumptions, the average photon index is tightly
constrained to be in the range $1.8<\langle\Gamma\rangle<2.0$. As a
first step we checked the effects of a different value for the average
photon index. Since in soft X--ray surveys an average slope of
$\langle\Gamma\rangle\sim 2$ is observed, we considered a distribution
centered at $\langle\Gamma\rangle=2$ (still having dispersion
0.2). Since the average unobscured AGN spectrum is now softer, the
flux ratio between the 2--10 keV and the 0.5--2 keV band is lower and,
as a consequence, more obscured AGN have to be added to match the hard
XLF of Ueda et al. (2003) and La Franca et al. (2005). In particular,
the $R_S$ and $R_Q$ values have to be increased from 4 to 5 and from 1
to 2, respectively. This larger number of obscured Compton-thin
sources, however, violates the limits imposed by the observed number
counts in the 0.5--2 keV band. Indeed, beyond the limits of shallow
soft X--ray surveys, these obscured sources emerge at faint fluxes due
to the favorable K--correction overproducing the observed counts. On
the contrary, a spectral distribution centered at
$\langle\Gamma\rangle=1.8$ implies a lower number of obscured AGN
required to match the hard XLF and the XRB spectrum, and turns out to
be in agreement with the source counts in all the considered bands. In
particular, as shown in Fig.~\ref{vhard}, assuming
$\langle\Gamma\rangle=1.8$ would help in matching the 5--10 keV counts
as observed in the HELLAS and CDFS surveys, which are otherwise
slightly underestimated when $\langle\Gamma\rangle=1.9$.  While
improving the fit of the very hard counts, the choice of an harder
($\langle\Gamma\rangle=1.8$) spectrum brings a drawback: the average
spectrum at various limiting fluxes does not match the observed one
anymore (e.g. Mateos et al. 2005). One may then wonder whether the
model should better reproduce the very hard 5--10 keV counts or the
average spectra at faint X--ray fluxes. In this respect we note that
the source detection in the 5--10 keV band and the corresponding
logN--logS determination suffer from systematic calibration
uncertainties much larger than those affecting the broader and softer
0.5--2 keV and 2--10 keV bands, mainly because the XMM-{\it Newton},
{\it Chandra} and BeppoSAX effective areas above 5 keV are rapidly
decreasing. It is also worth mentioning that keeping the dispersion at
$\sigma_{\Gamma}$=0.2 the harder average spectrum implies that the 30
keV XRB peak intensity (as measured by HEAO1) can be successfully
reproduced without including any contribution from Compton-thick
AGN. On the basis of what discussed above we feel that the slight
underprediction of the very hard counts in the baseline model with
$\langle\Gamma\rangle=1.9$ and $\sigma_{\Gamma}$=0.2 should not be
considered a major inconsistency.

Then the effects of varying the width $\sigma_{\Gamma}$ of the
spectral distribution were considered.  If we assume a very narrow
distribution, for instance $\sigma_{\Gamma}$=0, corresponding to a
single $\Gamma=1.9$ photon index, the model predictions for the 0.5--2
keV and 2--10 keV observational constraints are essentially the same
as those of the baseline model m2. Indeed, as shown in
Figs.~\ref{xdisp} and \ref{xrbdisp}, the effects introduced by a
distribution of spectral slopes increase towards high energies: the
wider is the slope distribution the larger is the contribution to the
XRB at say 30 keV (see Fig.~\ref{xrbdisp}). Therefore, the spectral
dispersion $\sigma_{\Gamma}$ and the space density of Compton-thick
sources are tightly correlated quantities. Assuming a distribution
with $\sigma_{\Gamma}$=0, the contribution of Compton-thin sources to
the 30 keV is lower than in our baseline model m2 and a larger number
of Compton-thick sources has to be added to match the XRB broad band
spectrum. More specifically one has to add twice as many Compton-thick
sources as in model m2. This possibility appears to already
overestimate the number of Compton-thick objects in the first BAT and
INTEGRAL AGN surveys above 20 keV.

The high energy cut--off in the AGN spectra is very poorly constrained
by present observations. The choice of $E_c$=200 keV in the adopted
spectral templates for both unobscured and obscured AGN is basically
driven by the intensity and the shape of the XRB spectrum above the
peak, which cannot be exceeded. Assuming a cut off energy $E_c=300$
keV, leaving the mean and dispersion of the photon index distribution
unchanged (1.9 and 0.2, respectively), the contribution of unobscured
and Compton-thin AGN saturate the XRB emission at 100 keV but
underestimate the 30 keV peak by about 20\%.  When trying to add
Compton-thick sources to fit the 30 keV emission, the XRB at 100 keV
is then overestimated. A global fit to the XRB spectrum using
$E_c=300$ keV can still be achieved but, in order to reduce the model
prediction at 100 keV with respect to that at 30 keV, one has to
assume a null dispersion in the photon index distribution, which
appears at variance with what discussed above.  It is concluded that
the average cut-off energy cannot be much higher than $E_c=200$ keV
especially if AGN X--ray spectral slopes are characterized by some
intrinsic dispersion.  More detailed considerations on the average
value of $E_c$ and the XRB constraints will be subject of future work
and will not be addressed here.

Very recently Hao et al. (2005) derived the optical luminosity
function of local low luminosity Seyferts, showing that their number
density is still increasing down to absolute magnitudes $M_B \simeq$
--16 to --14, which roughly correspond to log$L_x\lesssim 41$ assuming
standard AGN spectral energy distributions (e.g. Vignali et
al. 2005). We then explored the effects of extrapolating the HMS05 XLF
faint end slope down to log$L_x=41$, i.e. below the observational
limit of log$L_x=42$ which has been assumed in our calculations. The
predicted 0.5--2 keV counts of unabsorbed AGN exceed those observed by
HMS05 for type-1 AGN at faint fluxes ($<10^{-16}$ cgs). This suggests
that at low luminosities the slope of the AGN XLF should be flatter
than that observed at log$L_X>42$.

\subsection{The evolution of the fraction of obscured AGN with 
luminosity and redshift}

The evidence for a decrease in the fraction of obscured objects at
high X--ray luminosities is rapidly growing (Ueda et al. 2003,
Hasinger 2004, La Franca et al. 2005, Akylas et al. 2006), confirming
the early suggestion of Lawrence \& Elvis (1982). Further evidences
supporting a decreasing fraction of obscured AGN towards high
luminosities is provided by the comparison between the soft and hard
X--ray luminosity functions. In Section~\ref{ncnd} we showed that the
m2 model predictions, once folded with the appropriate sensitivity
curves, nicely fit the observed obscured to unobscured ratio in
different surveys. Any selection effect that could reconcile a
constant intrinsic ratio (i.e. Treister et al. 2004) is therefore
ruled out.

In model m2 the ratio $R$ between Compton-thin and unobscured AGN is
constrained to be in the range 2.6--4.8 at low luminosities ($R\sim
R_S$ for log$L_x\sim 42$) and in the range 0.6--1.5 at high
luminosities ($R \sim R_Q$ for log$L_x\sim 45$).  The population of
obscured quasars has then to be significantly smaller than that of
obscured Seyferts. However, it is worth mentioning that a model
without obscured QSOs at all ($R_Q=0$) is ruled out by the XLF
comparison (see Section~\ref{xlfsect}).

Additional evidences of a luminosity dependent ratio between obscured
type-2 AGN and unobscured broad line AGN were discussed by Barger et
al. (2005). We note that their findings, at the face value, imply
extremely high(low) ratios at low(high) luminosities.  As an example,
in the redshift range $z$=0.8--1.2 and at log$L_x$=42, the space
density of type-2 AGN is about 300 times higher than that of type-1
AGN.  These extremely high values for the type-2 to type-1 ratio have
been folded in some recent XRB models (Treister \& Urry 2005;
Ballantyne et al. 2006). More specifically Treister \& Urry (2005)
assume that {\rm all} the AGN are obscured at log$L_x$=42, while {\rm
none} is obscured at log$L_x$=46 clearly at variance with our
findings.  While the total (including both type-1 and type-2 objects)
AGN XLF obtained by Barger et al. (2005) is in good agreement with
those measured by Ueda et al. (2003) and La Franca et al. (2005),
their type-1 XLF appears to be in contrast with that measured by HMS05
especially at low luminosities. We believe that such a discrepancy is
due to some unaccounted for selection effect which makes the Barger et
al. (2005) type-1 AGN sample incomplete. Indeed only objects with a
significant detection of broad emission lines are considered as type-1
AGN by Barger et al. (2005).  It may well be that low luminosity,
broad line AGN with faint emission lines are misidentified in low
signal to noise optical spectra and thus not included in the type-1
sample.  The extremely high space density of objects classified as
optically normal galaxies by Barger et al. (2005) would support this
hypothesis. Indeed, the space density of optically normal galaxies
appear higher than that of AGN up to log$L_x$=43.5, again suggesting
that several X--ray sources classified as ``normal galaxies" may host
an AGN. In order to overcome such identification problems, HMS05 used
the X-ray plus optical classification scheme developed by Szokoly et
al. (2004) to efficiently identify AGN even with low signal to noise
optical spectra.  Not surprisingly the HMS05 type-1 AGN XLF at
log$L_x$=42 is about two orders of magnitude higher than that of
Barger et al. (2005). We refer to Szokoly et al. (2004) for a detailed
discussion on the incompleteness affecting type-1 AGN samples selected
only by means of moderate quality optical spectroscopy.

The dependence of the obscured AGN fraction with redshift is still 
debated. Synthesis models based on pre-{\it Chandra} and XMM results 
(e.g. Gilli et al. 2001) were favoring an increasing fraction 
of obscured sources with redshift.
This hypothesis was also put forward by Fabian (1999), who postulated the
existence of a large number of obscured QSOs at high redshift. 
More recently, some evidence for an increasing fraction towards
high--z was discussed by La Franca et al. (2005), while other authors
(Ueda et al. 2003, Akylas et al. 2006) did not find a statistically
significant variation of obscured sources with redshift. In particular
Akylas et al. (2006) suggested that the apparent increase in the
obscured AGN fraction with redshift is due to a systematic
overestimate of the column densities measured in high redshift sources
where the absorption cut--off is shifted towards low energies (see
also Tozzi et al. 2006). The comparison between the soft and hard XLF
presented in the previous Sections leaves little room for any
significant evolution of the obscured AGN fraction with
redshift. Similar results are also obtained by Hasinger (2006).
We note however that at high redshift, where the absorption cut-off is
redshifted out of the soft band, some mildly obscured AGN may
incorrectly be classified as unobscured and included in the soft XLF.
The obscured to unobscured AGN ratio $R$ could then be biased towards
lower values and any weak redshift dependence of the obscured AGN
fraction would then be missed by our approach. Determining the exact
behaviour of $R$ with redshift would require an extensive
investigation of the biases affecting the various estimates appeared
in the literature, which is beyond the scope of this paper. For the
time being we can safely state that the strongest observed dependence
of the obscured AGN fraction is on luminosity. We stress that the
number of high redshift obscured QSOs predicted by the current
modeling appears to be in good agreement with recent estimates based
on mid to near infrared (Spitzer) selection (Martinez-Sansigre et
al. 2005), which should efficiently detect even heavily obscured
objects. The intrinsic ratio between obscured and unobscured QSOs at
$z\sim 2$ as measured by Spitzer is in the range $\sim 1-3$
(Martinez-Sansigre et al. 2005,2006). The corresponding best fit ratio
of model m2 is 1 when only Compton-thin obscured QSOs are considered
and 2 when Compton-thick QSOs are included.

\subsection{The XRB normalization}\label{xrbnorm}

As stated in the previous Sections, we did not consider the XRB
intensity as a primary constraint, rather we make an extensive use of
the soft and hard XLF and source counts to get a complete census of
moderately obscured AGN and then add Compton-thick sources to fit the
XRB flux as measured by HEAO1.  Since all the recent XRB measurements
performed by imaging instruments at $E<10$ keV appear to have the same
slope as the one measured by HEAO-1 but higher normalization, it has
become a common practice in recent literature (Ueda et al. 2003,
Treister \& Urry 2005) to renormalize upward by $\sim 30-40\%$ the
entire broad band 3--400 keV XRB spectrum measured by HEAO-1 and then
fit this higher background. In our view, this renormalization it is
not fully justified. Even assuming that a calibration problem occurred
in the low energy (3--40 keV) A2 experiment on board HEAO-1, this does
not necessarily imply that the 15--100 keV XRB measurements by the A4
detector is also wrong (incidentally, the A4 flux appears to be higher
by $\sim 5-10\%$ than that of A2 in the overlapping energy
range). The XRB flux at 30 keV measured by HEAO-1, has been
indeed recently confirmed (within the statistical errors) by INTEGRAL
(Churazov et al. 2006) and BeppoSAX/PDS (Frontera et al. 2006). It is
also important to point out that in order to match the 1--10 keV XRB
flux as measured by BeppoSAX and XMM the XLF should be integrated down
to very low luminosities (log$L_X \simeq$ 40). Such a solution is
ruled out by the limits imposed by the source counts (see
Section~\ref{explo}).
A similar problem is present among those AGN synthesis models which
rescale upwards by a factor 1.3 or 1.4 the entire HEAO-1 XRB spectrum
(e.g. Treister \& Urry 2005, Ballantyne et al. 2006). Indeed their
predicted 0.5--2 keV and 2--10 keV logN-logS are largely inconsistent
with the observations.

The total XRB flux estimated by Hickox \& Markevitch (2006) combining
an estimate of the ``unresolved" background in the {\it Chandra} deep
fields with that already resolved by surveys at brigther fluxes,
settles in between the original HEAO1 measure and the BeppoSAX and XMM
fluxes. In both the 0.5--2 keV and 2--8 keV bands the unresolved
intensities cannot be explained by extrapolating to zero fluxes the
observed logN--logS. This is especially true in the soft (0.5--2 keV)
band suggesting a steepening of the X--ray counts below the present
limits most likely due to ``normal" and star forming galaxies (see
Fig.\ref{soft}) as proposed by Ranalli et al. (2003) and Persic \&
Rephaeli (2003).  At the face value the above described findings
indicate that AGN synthesis models should not saturate the entire
background flux around few keV in order to leave some room for the
contribution of non-AGN sources. By fitting the HEAO1 background above
10 keV (see Fig.~\ref{xrb}) our model predictions fall short the
Hickox \& Markevitch (2006) level by $\sim 20\%$ in the 1--2 keV band
and thus the contribution of starforming galaxies could be easily
accomodated.  A higher intensity of the XRB around the 30 keV peak can
be accounted for by increasing the contribution of Compton-thick
AGN. More specifically should the high energy XRB be higher by $\sim
30-40\%$, the number density of Compton-thick AGN would increase by a
factor of 2-3.


\subsection{The number density of Compton-thick AGN}

Since the cosmological properties of the Compton-thick AGN population
are unknown, we had to rely on a few minimal assumptions in our
approach.  The first is that the XLF of Compton-thick AGN is the same
as that of moderately obscured AGN. This for instance implies that the
number of Compton-thick Seyfert 2s is similar to the number of
obscured Compton-thin Seyfert 2s, in agreement with the observations
in the local Universe (e.g. Risaliti et al. 1999), but also that the
number of Compton-thick QSOs is similar to that of moderately obscured
QSOs. Given the very small statistics on Compton-thick QSOs (Comastri
2004), their absolute space density remains unknown and cannot be
efficiently constrained by the current modeling. 

Another underlying assumption is that the number density of heavily
Compton-thick AGN (log$N_H >$ 25) is the same as that of mildly
Compton-thick AGN (24 $< {\rm log}N_H <$ 25). Since their spectrum is
Compton down-scattered up to several hundreds of keV the space density
of heavily Compton-thick AGN is essentially unconstrained by the
currently available data. Had we neglected the contribution of heavily
Compton-thick AGN the integrated XRB emission would decreases by only
5\% at 30 keV, without modifying the predicted fraction of Compton
Thick AGN which successfully match INTEGRAL, Swift and CDFS
observations.

Finally, the estimated number density of Compton-thick AGN, especially
that of heavily Compton-thick objects, critically depends on the
normalization assumed for their pure reflection spectrum with respect
to the intrinsic nuclear emission. In Section~\ref{abs} the reflected
spectrum was normalized such as to produce 2\% of the 2-10 keV
intrinsic emission, but the actual average value is highly
uncertain. A reflected fraction of 1\% or less, which is still in
agreement with the observed X-ray spectra of Compton-thick AGN, would
imply a number of heavily Compton-thick objects a factor of $>2$
higher than previously assumed to produce the same contribution to the
high-energy XRB spectrum. It is however worth noting that the number
of Compton-thick AGN cannot be increased arbitrarily without 
violating the limits imposed by the local black hole mass density  
(e.g. Marconi et al. 2004). A quantitative investigation is beyond 
the purposes of this paper and will be subject of future work.


\section{Conclusions and future work} \label{conclusions}


The most important results obtained in this work can be summarized 
as follows:

1) The ratio $R$ between moderately obscured and unobscured AGN is
found to decrease from about 4 at log$L_x\lesssim 42$ to about 1 at
log$L_x\gtrsim 45$. A constant $R$ value is ruled out by: i) the
comparison between the soft and hard XLF and ii) the decreasing
fraction of obscured sources as a function of luminosity in different
surveys, which cannot be accounted for by selection effects. The XLF
comparison leaves little room for any significant increase of the
obscured AGN fraction towards high redshifts.

2) Although the fraction of obscured AGN is found to decrease with
luminosity a non-negligible population of obscured QSOs is still
required. In particular, the ratio between moderately
obscured and unobscured QSOs is constrained to be in the range
0.6--1.5.


3) An intrinsic distribution in the AGN photon indices centered at
$\langle\Gamma\rangle=1.9$ and with dispersion $\sigma_{\Gamma}$=0.2
has been assumed in our baseline model. The spectral dispersion
enhances by 20--30\% the contribution of Compton-thin AGN to the 30
keV XRB peak with respect to a single slope average spectrum, but
still is not sufficient to match the XRB intensity, calling for a
significant contribution from Compton-thick AGN.

4) A sizable population of Compton-thick AGN has to be assumed to match
the high energy XRB spectrum as measured by HEAO-1. Their number
density is estimated to be of the same order of that of moderately
obscured AGN. The model predictions are in extremely good agreement
with the fraction of Compton-thick objects observed in the Chandra
Deep Field South and in the first Swift and INTEGRAL catalogs of AGN
selected above 10 keV.

5) A detailed investigation of the parameter space suggests that the
"best fit" parameters of the baseline model are relatively well
constrained in a global sense. This means that a small variation of a
single parameter may easily violate one or more of the observational
constraints. The model makes also quantitative predictions on the AGN
properties beyond the present limits and especially above 10 keV.


6) The XRB intensity in the 2--10 keV energy range as measured by
imaging detectors, cannot be easily accounted for by our model without
violating a number of well constrained observational facts such as the
source counts and the average spectra of X--ray selected AGN at
various limiting fluxes. A higher normalization of the $<$ 10 keV XRB
spectrum can still be reconciled with our findings assuming that
non--AGN sources contribute to $\sim 10-20\%$ of the XRB around a few
keV. Such a possibility appears to be supported by independent
analyses. The issue of the XRB normalization above 10 keV is tightly
linked with the space density of Compton-thick AGN. The higher is the
30 keV peak intensity the larger the number of Compton-thick sources.

~

The approach discussed in this paper will constitute a solid reference
framework which we will use for future investigations of the obscured
AGN population. In particular the model predicted space density of
Compton-thin and Compton-thick AGN will be compared with that obtained
by infrared Spitzer surveys (i.e. Polletta et al. 2006). Also, the
predictions about the AGN content of the X--ray sky at energies above
20 keV, especially for what concerns the Compton-thick AGN population,
will be tested by the high-energy imaging instruments on board planned
and/or proposed missions such as Symbol--X and NeXT.

\acknowledgements

RG and AC acknowledge support from the Italian Space Agency (ASI)
under the contract ASI--INAF I/023/05/0. We thank Piero Ranalli for
providing the numerical code to compute Compton-thick spectra and
Marcella Brusa for help with Fig.~\ref{picoplot}. We also thank Mike
Revnivtsev, Yoshihiro Ueda and Fabio La Franca who provided data from
their papers in a machine readable format. We thank Giancarlo Setti,
Gianni Zamorani, Marco Salvati, Alessandro Marconi, Guido Risaliti,
Fabrizio Fiore, Pierluigi Monaco and Stefano Bianchi for useful
discussions.

\end{document}